\documentclass[useAMS,usenatbib]{mn2e}
\usepackage{graphicx}
\usepackage{amssymb}
\usepackage{amsmath}
\usepackage{caption}
\usepackage{subcaption}
\usepackage{epstopdf}

\title[CO mass upper limits and excitation in the Fomalhaut ring]{CO mass upper limits in the Fomalhaut ring - the importance of NLTE excitation in debris discs and future prospects with ALMA}
\author[L. Matr\`a et al.]{L. Matr\`a$^{1,2}$\thanks{E-mail:
l.matra@ast.cam.ac.uk}, O. Pani\'c$^{1}$, M. C. Wyatt$^{1}$, W. R. F. Dent$^{3}$ \\
\ \\
$^{1}$Institute of Astronomy, University of Cambridge, Madingley Road, Cambridge CB3 0HA, UK \\
$^{2}$European Southern Observatory, Alonso de C\'ordova 3107, Vitacura, Santiago, Chile \\
$^{3}$ALMA SCO, Alonso de C\'ordova 3107, Vitacura, Santiago, Chile
}
\begin{document}

\date{Accepted 2014 December 8}

\pagerange{\pageref{firstpage}--\pageref{lastpage}} \pubyear{2002}

\maketitle

\label{firstpage}

\begin{abstract}
In recent years, gas has been observed in an increasing number of debris discs, though its nature remains to be determined. Here, we analyse CO molecular excitation in optically thin debris discs, and search ALMA Cycle-0 data for CO J=3-2 emission in the Fomalhaut ring. No significant line emission is observed; we set a 3-$\sigma$ upper limit on the integrated line flux of 0.16 Jy km s$^{-1}$. We show a significant dependency of the CO excitation on the density of collisional partners $n$, on the gas kinetic temperature $T_k$ and on the ambient radiation field $J$, suggesting that assumptions widely used for protoplanetary discs (e.g. LTE) do not necessarily apply to their low density debris counterparts. When applied to the Fomalhaut ring, we consider a primordial origin scenario where H$_2$ dominates collisional excitation of CO, and a secondary origin scenario dominated by e$^-$ and H$_2$O. In either scenario, we obtain a strict upper limit on the CO mass of 4.9 $\times$ 10$^{-4}$ M$_{\oplus}$. This arises in the non-LTE regime, where the excitation of the molecule is determined solely by the well-known radiation field. In the secondary scenario, assuming any CO present to be in steady state allows us to set an upper limit of $\sim$55\% on the CO/H$_2$O ice ratio in the parent planetesimals. This could drop to $\sim$3\% if LTE applies, covering the range observed in Solar System comets (0.4-30\%). Finally, in light of our analysis, we present prospects for CO detection and characterisation in debris discs with ALMA.

\end{abstract}


\section{Introduction}

Observations show many young pre-main sequence stars to be surrounded by discs of gas and dust. These are known as protoplanetary discs, since it is within them that the planet formation process is inferred to begin. As the star evolves to the main sequence, the products of the planet formation process appear: not only planets themselves, but also debris, in the form of asteroids, comets, and dust. In exosolar systems, this debris takes the form of rings and broader belts, and can carry important signatures of both the presence and the formation of planets \citep[see review by][and references therein]{Wyatt2008}.

In the past, these systems have been extensively studied as dusty environments, since most of the gas is expected to dissipate in the transition between the protoplanetary and debris stage of disc evolution. However, in a few bright systems such as 49 Ceti and $\beta$ Pictoris, non-negligible amounts of gas have been known to exist for decades \citep[][]{Zuckerman1995,Slettebak1975}. 
In recent years, telescopes of increased sensitivity have been providing new discoveries of gas emission in debris discs \citep[see review by][and references therein]{Matthews2014}, particularly in young $\lesssim$50 Myr systems \citep[e.g.][]{Moor2011, RiviereMarichalar2012, RiviereMarichalar2014}. Notably, ALMA is now also able to provide detailed maps of the location and dynamics of the gas, especially through CO millimetre and sub-millimetre emission lines, giving us the potential to better constrain its origin. This has been already done for the $\beta$ Pictoris \citep{Dent2014} and HD 21997 \citep{Kospal2013} systems. For $\beta$ Pictoris, the CO distribution in the disc suggests its origin to be destructive collisions of icy comet-like bodies \citep[as also suggested for the CO around 49 Ceti,][]{Zuckerman2012}, while for HD 21997 it is proposed that the CO is of primordial origin. However, very little is actually known about the gaseous components of debris discs. 

One of the clearest examples of dusty debris environments is found in the nearby, $\sim$440 Myr old \citep{Mamajek2012} Fomalhaut system. The star harbours a debris ring which has been imaged in scattered light at a semi-major axis of $\sim$140 AU \citep[][]{Kalas2005}. The presence of a planet (Fomalhaut b) was first inferred from the eccentricity and sharply truncated inner edge of the debris ring, and subsequently confirmed through direct imaging \citep{Kalas2008}. However, follow-up studies show its orbit to be highly eccentric, and question the object's nature, origin and interaction with the dust ring \citep[e.g.][]{Kalas2013,Beust2014, Tamayo2014}. Traces of gas are yet to be found in the system.

A high-resolution 850 $\mu$m ALMA Cycle-0 continuum image of the north-west side of the ring was obtained by \citet{Boley2012}, and indicated that the population of planetesimals is concentrated in a sharp-edged $\sim$16 AU-wide belt. In this work, we retrieve these data to search for CO J=3-2 line emission within the belt. We also investigate the dependence of the excitation state of CO on local physical parameters, and show how this affects estimation of the limit on the total mass of CO in debris systems like Fomalhaut. 

The paper is structured in the following way: Sect. 2 describes the ALMA data and the results of our search. In Sect. 3, we describe how the excitation state of the CO molecule and hence total CO mass estimates depend on local physical quantities, and in Sect. 4 we show how our CO analysis can be applied through the example of the Fomalhaut ring. Finally, Sect. 5 consists of a discussion of the significance of our results for Fomalhaut and for gas-bearing debris systems, with a particular focus on future ALMA observations.

\section[]{Observations and Results}

The Fomalhaut ring was observed in 2011 by ALMA (with pointing centred on the position of Fomalhaut b, RA: $22^h57^m38^s.65$, Dec: $-29^d37\arcmin12\farcs6$) for 140 minutes (Band 7). A continuum sensitivity of $\sim$60 $\mu$Jy beam$^{-1}$ was achieved for a beam of size 1$\farcs$5 $\times$ 1$\farcs$2, with upper and lower sidebands at frequencies of 357 and 345 GHz, respectively. The flux calibration uncertainty was estimated to be $\sim$10\%.  A more detailed description of the observations and data reduction can be found in \citet{Boley2012}.
Here we repeated the imaging section of the data reduction using the CASA software v.4.1.0 \citep{McMullin2007} in order to search for CO J=3-2 line emission. In particular, we retained the channel width of the calibrated data (488.28 kHz, i.e. 0.424 km/s) rather than combining channels together as was originally done to obtain a continuum image. We selected a 200-channels wide (85 km/s) spectral window around the CO line, expected at 345.789 GHz \citep[v$_{hel}$=6.5 $\pm$ 0.5 km/s,][]{Gontcharov2006}. 

\begin{figure}
\vspace{-4mm}
 \hspace{-4mm}
  \includegraphics*[scale=0.35]{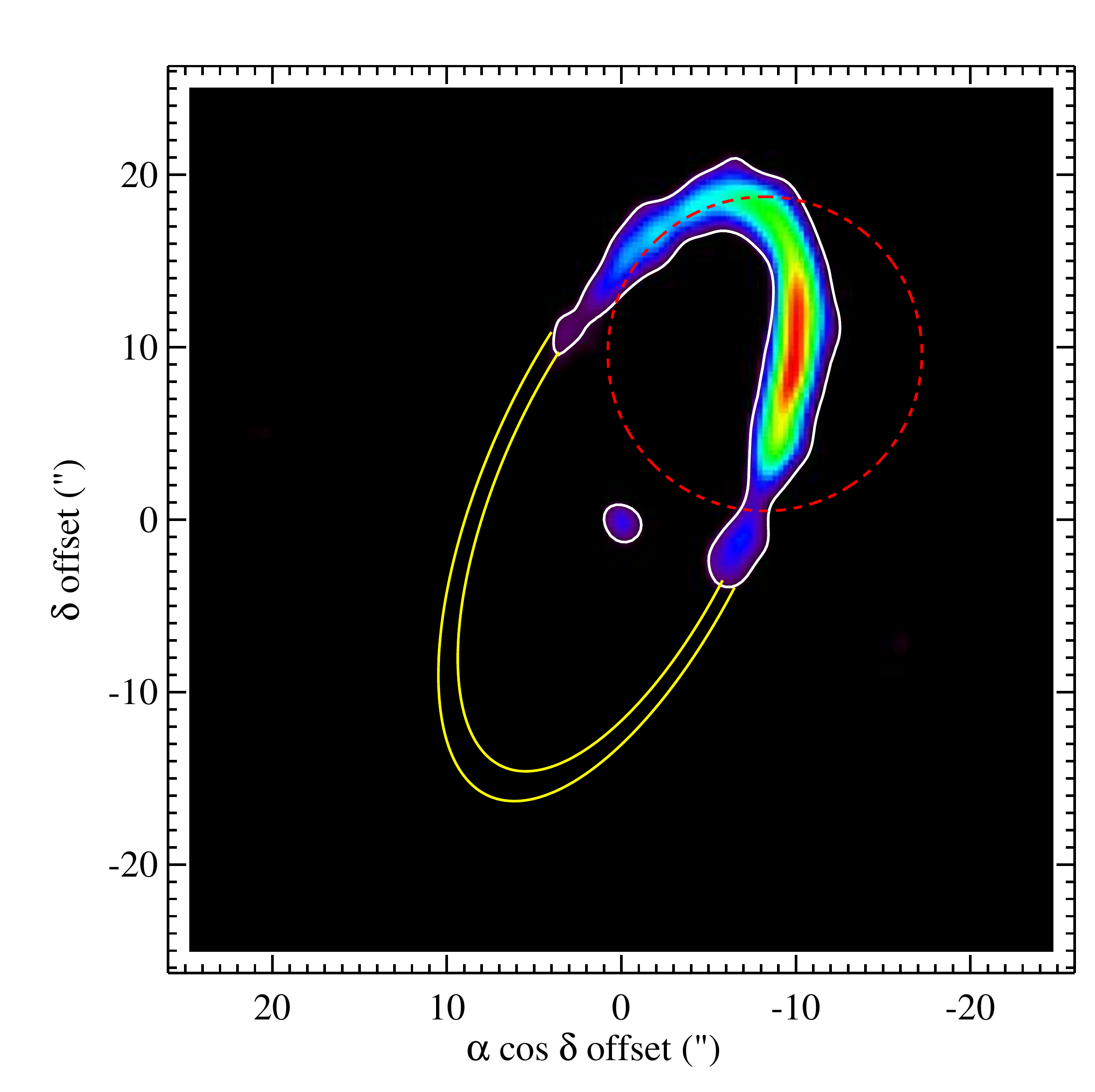}
\vspace{-7.5 mm}
  \caption{Continuum non-primary-beam-corrected image of the Fomalhaut ring. The white contour is at the 4-$\sigma$ emission level, representing the pixels selected in our analysis (where we later exclude those associated with the location of the star). The dashed red circle represents the primary beam, whereas the yellow lines are the best-fit inner and outer radii of the ring from \citet{Boley2012}.}
\label{fig:1}
\end{figure}

We also recovered the non-primary-beam-corrected continuum image from \citet{Boley2012}. We used this to select pixels where continuum emission observed is $\ge$ 4$\sigma$, and subsequently discarded pixels associated with the star (Fig.\ref{fig:1}).
Assuming the ring to be in Keplerian rotation \citep[where M$_{\ast}$=1.92 M$_{\odot}$,][]{Mamajek2012} and using orbital parameters from \citet{Kalas2005}, we assign to each pixel a spectral channel corresponding to its projected velocity vector. The expected observed velocities lie in the ranges between +0.5 and -3.0 km/s or between -0.5 and +3.0 km/s, where the degeneracy is due to the direction of rotation with respect to the line of sight. Taking this into account, we expect any CO emission to fall within 14 channels around the J=3-2 transition frequency (corrected for the velocity of the star). First, we visually inspect the reduced data cube channel by channel, and produce a moment-0 map by co-adding images in the expected pixel velocity ranges. We do not detect any significant emission anywhere in the map.

\begin{figure}
 \vspace{-3.5mm}
 \hspace{-6mm}
  \includegraphics*[scale=0.45]{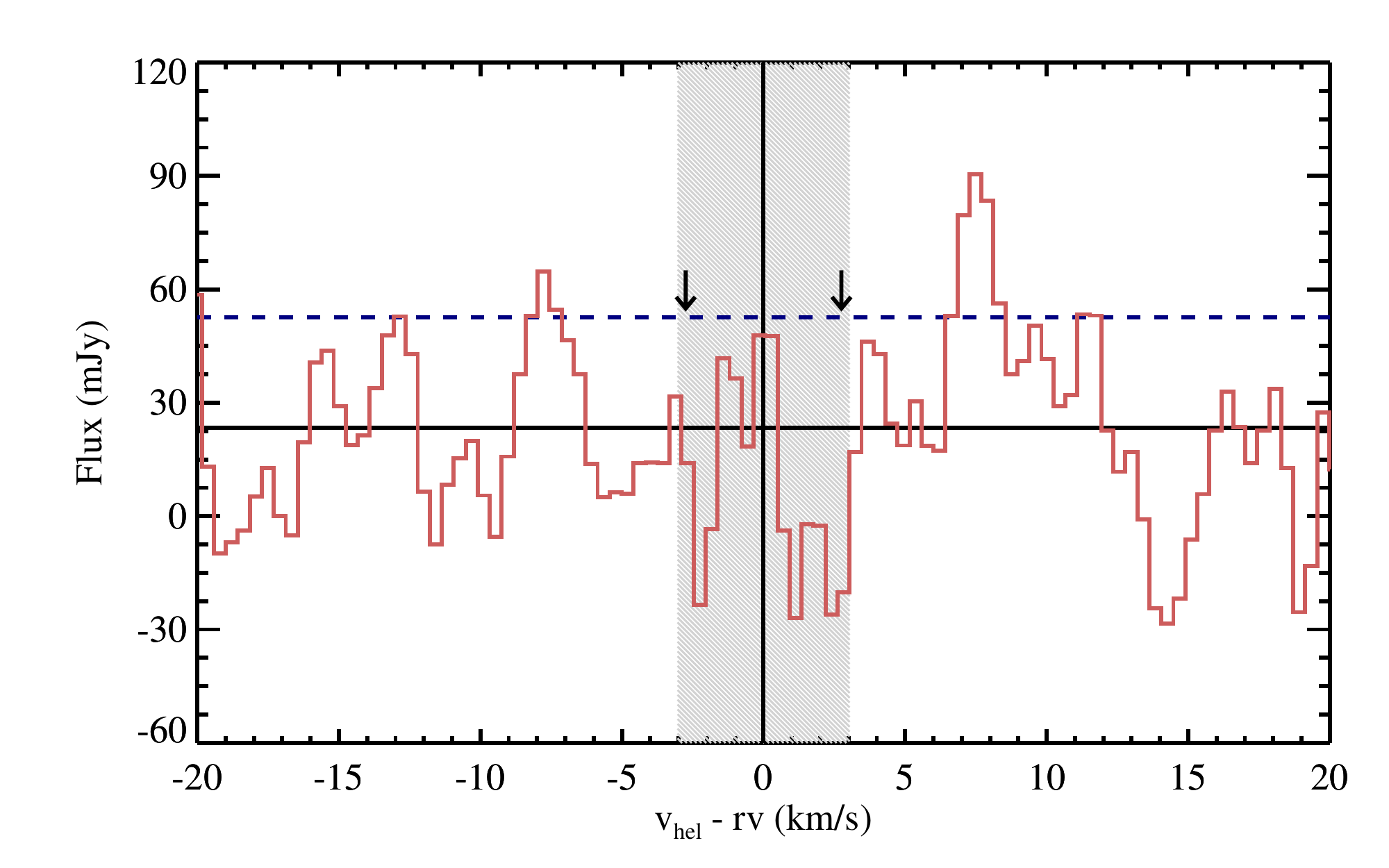}
\vspace{-5 mm}
  \caption{Spectral emission integrated over the selected ring area (see Fig. 1). The horizontal black and dashed blue lines represent the mean and the 1-$\sigma$ level of the spectrum, respectively. The shaded area represents the velocity range where we would expect to see CO emission from the ring; the black arrows indicate the expected locations of the line peak, for the two possible signs of the inclination angle.}
\label{fig:2}
\end{figure}

To improve our chance of detection and assuming that any CO, if present, will be co-located with the dust component, we integrate the observed emission over the selected ring area. At first we do not take into account the velocity assigned to each pixel; the resulting spectrum is shown in Fig. \ref{fig:2}. This procedure eliminates the noise component originating from pixels where no emission is expected. No CO is detected within the expected pixel velocity range. The $\sim$2$\sigma$ peak observed at 7.5 km/s is unlikely to be real, due to the low significance and inconsistency with the ring velocities. We note that the error on the Fomalhaut radial velocity measurement is relatively small ($\pm$0.5 km/s). In addition, the mean continuum level of the spectrum (23 mJy) is in line with both the ALMA continuum flux estimate in the observed region \citep[20.5 mJy;][]{Boley2012} and with previous single-dish observations \citep[81.0$\pm$7.2 and 97$\pm$5 mJy, ][]{Holland1998,Holland2003}, if we consider that the selected area accounts for only part of the full ring emission, and that the sensitivity drops considerably at the edges of the primary beam (i.e. near the semi-minor axis of the ring). 

We then improve our detection level further: for the selected pixels, we only sum fluxes in the spectral channels identified for the velocity of the ring at that location. Using this method, we obtain fluxes of $22\pm30$ mJy and $2\pm30$ mJy (depending on the degeneracy in the sign of the inclination mentioned above), where 30 mJy represents the 1-$\sigma$ error on the spatially integrated flux in a single frequency channel, as shown in Fig.\ref{fig:2}. 
These values are well below the 3-$\sigma$ threshold of 90 mJy; we therefore conclude that no CO is detected within the Fomalhaut ring.

We then proceed to calculate upper limits.  
For a conservative estimate, we add a 10\% absolute flux calibration error to our 3-$\sigma$ uncertainty on the spatially integrated flux (90 mJy per spectral channel). This leads to a 3-$\sigma$ upper limit of 4.2 $\times$ 10$^{-2}$ Jy km s$^{-1}$ (4.8 $\times$ 10$^{-22}$ W m$^{-2}$) on our spatially integrated flux within a single spectral channel (calculated using our channel width of 0.424 km s$^{-1}$). 
We finally note that the selected area is only part of the whole ring, and hence rescale our result using the ratio of continuum flux of the whole ring \citep[as measured by][]{Boley2012} over that of the partial ring (85 mJy/23 mJy), under the assumption that the CO and sub-mm dust emission are co-located and axisymmetric. This yields a final 3-$\sigma$ CO J=3-2 line flux upper limit of 0.16 Jy km s$^{-1}$ (1.8 $\times$ 10$^{-21}$ W m$^{-2}$).

\section[]{Excitation of CO rotational levels}
\label{Sect:molexc}
If CO line emission observed is optically thin, the integrated line flux can be used to derive the total mass of CO in the system. In fact, the integrated line luminosity $L_{j,j-1}$ of an optically thin line can be obtained through
\begin{equation} 
L_{j,j-1}=h\nu_{j,j-1}A_{j,j-1}N_j,
\end{equation}
where $N_j$ is the number of CO molecules in rotational level $j$ (the \textit{level population}), $\nu_{j,j-1}$ is the line rest frequency, $h$ is Planck's constant, and $A_{j,j-1}$ is the Einstein $A$ coefficient for the transition, representing the probability per second that a molecule in level $j$ will decay to level $j-1$. The integrated line flux $F_{j,j-1}$ can then be linked directly to the total CO mass
\begin{equation} 
M=\frac{4\pi m d^2}{h\nu_{j,j-1}A_{j,j-1}}\frac{F_{j,j-1}}{x_j},
\end{equation}
where $m$ is the mass of the molecule, $d$ is the distance to the system from the observer, and $x_j$ is the fraction of molecules that are populating the upper level of the transition ($N_j$/$N_{tot}$, the \textit{fractional level population}). All the parameters needed for the calculation are known except for the upper level fractional population $x_j$, which is dependent on how different excitation mechanisms excite the CO molecule.

It is instructive at first to simplify the problem by considering a two-level system, where $u$ and $l$ indicate the upper and lower level of the transition.  The level populations $N$ can be derived from statistical equilibrium, including both radiative and collisional terms, using
\begin{equation} 
N_ln\gamma_{l,u}+N_lB_{l,u}J_{l,u}=N_un\gamma_{u,l}+N_uB_{u,l}J_{u,l}+N_uA_{u,l}
\end{equation}
where $n$ is the density of the dominant collisional partner, $B$ is the Einstein $B$ coefficient (representing the probability of either a radiative absorption, $B_{l,u}$, or a stimulated radiative emission, $B_{u,l}$, per unit time and unit mean intensity) and $\gamma_{u,l}$ is the collisional rate coefficient for the transition. $J$ is the mean intensity of the radiation field at the transition frequency $\nu_{u,l}$, and is defined as
\begin{equation}
J_{u,l}=\frac{1}{4\pi}\int_{\Omega_s}I_{u,l} d\Omega,
\end{equation} 
where we are integrating the intensity $I_{u,l}$ over the solid angle $\Omega_s$ subtended by its source, and then dividing by $4\pi$ to obtain the mean intensity (i.e. averaged over the whole sky).

Eq. 3 shows that a transition occurs via a mixture of both collisional and radiative processes, namely collisional excitation and radiative absorption from the lower level (left hand side), and collisional de-excitation, stimulated emission and spontaneous emission from the upper level (right hand side). The Einstein coefficients can be related using
\begin{equation}
A_{u,l}=\frac{2h\nu_{u,l}^3}{c^2}B_{u,l}=\frac{2h\nu_{u,l}^3}{c^2}\frac{g_{l}}{g_{u}}B_{l,u},
\end{equation} 
where $g_u$ and $g_l$ are the degeneracies of each level. In addition, the collisional rate coefficients are linked via
\begin{equation}
\gamma_{l,u}=\frac{g_u}{g_{l}}\gamma_{u,l}e^{-h\nu_{u,l}/kT_k},
\end{equation}
where $T_k$ is the kinetic temperature of the gas.
Making use of Eq. 3, 5, 6, and rearranging yields
\begin{equation} 
\frac{N_u}{N_l}=\frac{\frac{g_u}{g_l}\left(\frac{c^2}{2h\nu_{u,l}^3}J_{u,l}\frac{n_{crit_{u,l}}}{n}+e^{-h\nu_{u,l}/kT_k}\right)}{1+\left(1+\frac{c^2}{2h\nu_{u,l}^3}J_{u,l}\right)\frac{n_{crit_{u,l}}}{n}},
\end{equation} 
where $n_{crit_{u,l}}$ is the critical density, defined as $n_{crit_{u,l}}$ = $A_{u,l}$/$\gamma_{u,l}$. 

We find that local thermodynamic equilibrium (LTE) is analytically recovered at high $n$, where collisions dominate the excitation and de-excitation process
\begin{equation} 
n\gg n_{crit_{u,l}} \Rightarrow \frac{N_u}{N_l}\approx\frac{g_u}{g_l}e^{-h\nu_{u,l}/kT_k}.
\end{equation} 
In this limit, the level populations follow the Boltzmann distribution, and the partition function $Z=\sum\limits_ig_ie^{-E_i/kT_k}$ can be used to calculate the fractional populations
\begin{equation}
n\gg n_{crit_{u,l}} \Rightarrow x_j=\frac{N_j}{N_{tot}}=\frac{g_j}{Z}e^{-E_j/kT_k},
\end{equation}
where $E_j$ represents the energy of level $j$ above the ground state, and other symbols have usual meanings. This also implies that if $n\gg n_{crit_{u,l}}$ the \textit{excitation temperature} $T_{ex}$ (defined through Eq. 8 for any two levels and in any excitation regime) is the same for all levels and corresponds to the kinetic temperature $T_k$. In other words, the levels are said to be \textit{thermalised}.

This is not applicable to the low $n$ limit, where radiative excitation and de-excitation, together with spontaneous emission, dominate over collisional processes
\begin{equation} 
n\ll n_{crit_{u,l}} \Rightarrow \frac{N_u}{N_l}\approx\frac{g_u}{g_l}\frac{\frac{c^2}{2h\nu_{u,l}^3}J_{u,l}}{1+\frac{c^2}{2h\nu_{u,l}^3}J_{u,l}}.
\end{equation}
In this case, the level populations do not follow the Boltzmann distribution, and hence we cannot make use of the partition function to obtain fractional populations $x_j$. In addition, the excitation temperature is different for each pair of levels and also different from the kinetic temperature. In other words, the excitation is said to be \textit{subthermal}.

This clearly highlights the importance of the density of collisional partners $n$ in determining whether the level populations are dominated by radiative or collisional processes, and hence in calculating the level populations $x_j$. In reality, there is a further complication given by the fact that a molecule is composed of a plethora of energy levels, implying that the level populations are, to different extents, coupled together. We here neglect the influence of transitions from excited vibrational and electronic states on the population of the rotational levels; this effect will be fully treated in upcoming work. If we approximate the molecule to have rotational levels up to $j_{max}$, the full statistical equilibrium equation for each level $j$ becomes
\begin{equation} 
\begin{split}
N_{j-1}B_{j-1,j}J_{j-1,j}+N_{j+1}B_{j+1,j}J_{j+1,j}+N_{j+1}A_{j+1,j}+ \\
+\sum\limits^{j_{max}}_{\substack{m=0 \\ m\neq j}}N_{m}n\gamma_{m,j}=N_{j}B_{j,j-1}J_{j,j-1}+N_{j}A_{j,j-1}+ \sum\limits^{j_{max}}_{\substack{p=0 \\ p\neq j}}N_{j}n\gamma_{j,p},
\end{split}
\end{equation}
where symbols have their usual meanings. Similarly to Eq. 3, the left hand side contains all terms that populate level $j$, namely absorption from level $j-1$, stimulated and spontaneous emission from level $j+1$, and collisional transitions from any other level. The right hand side, on the other hand, includes terms that depopulate level $j$, namely stimulated and spontaneous emission to level $j-1$, and collisional transitions to any other level. We therefore end up with $j_{max}$+1 linear equations in $j_{max}$+1 unknowns that need to be solved simultaneously. However, these equations are not linearly independent, so one of them has to be replaced by the normalisation condition
\begin{equation} 
N_{tot}=\sum\limits^{j_{max}}_{j=0}N_{j},
\end{equation}
where if we set $N_{tot}$ to 1, we can substitute all $N_{j}$ terms with $x_{j}$ (i.e., set $\sum\limits^{j_{max}}_{j=0}x_{j}=1$) and upon solution of the system of equations immediately recover the fractional population of each level.

\section{Application to the Fomalhaut ring}

Given our upper limit flux estimate from the ALMA observations, we use our CO excitation analysis to find the maximum amount of CO mass that could be present in the Fomalhaut ring, while still evading detection. 
In Sect. 3, we have shown that the result will depend on the fractional population of the upper level of the transition, which will in turn depend on physical quantities $n$, $T_k$ and $J_{j,j-1}$, whose values vary within the disc. Hence, a physical model of the disc including collisional partners, radiation sources and CO gas needs to be built for deriving the excitation state of the molecule locally. 

For simplicity, dust, collisional partners and CO are assumed to be co-located and to follow the same density distribution within the ring. This is a likely scenario if the origin of any gas present were to be associated with icy bodies within the ring, such as found in the $\beta$ Pic system \citep{Dent2014}. 
In our gas and dust models, we also assume the ring to be axisymmetric. This does not have a significant effect on our results, since the eccentricity is very low \citep[0.11, ][]{Kalas2005}, meaning that the difference in length between semi-major and semi-minor axes is negligible (0.6\%). For example, the flux originating from a dust particle located at semi-minor axis and being absorbed by a CO molecule located at semi-major axis will only be 0.6\% higher than if we considered both dust and CO to be on a circular orbit of radius equal to the semi-major axis. Furthermore, for the lowest CO rotational levels, we expect the most significant component of the flux impinging on a CO molecule in the Fomalhaut ring to originate from the cosmic microwave background (CMB), removing the dependence on the dust disc structure. (see Sect. 4.1 below).

\subsection{The radiation environment}
\label{sect:radenv}
The mean radiation field $J$ has three different contributions originating from the surrounding dust, the CMB and the star. For each transition, we have $J=J_{dust}+J_{cmb}+J_{star}$, where each component can be derived using Eq.~4. Due to its isotropic and optically thick nature, $J_{cmb}$=$B_{cmb}$, where $B_{cmb}$ is the Planck function of the CMB. The contribution from the star is given by $J_{star}$=$\frac{1}{4\pi}\frac{\pi R^2_{\ast}}{d^2}B_{star}$, where $R_{\ast}$ is the radius of the star, $d$ its distance from the CO molecule, and $B_{star}$ its Planck function. The $J_{dust}$ contribution is dependent on the radial and vertical distribution, on the temperature profile, and on the total mass of dust in the disc. Therefore, a dust model based on best-fit to sub-mm continuum observations is necessary to calculate $J_{dust}$ at each point in the disc. For the Fomalhaut ring, we use a dust model in agreement with the best-fit to the ALMA continuum image obtained by \citet{Boley2012}. This consists of a Gaussian radial profile with FWHM of 16 AU and centred at a radius of 143 AU, and an exponential vertical profile with an opening angle of 1$^{\circ}$. At each location within the CO model grid, we use line-of-sight integration to calculate dust continuum fluxes from the dust model through different pixels covering the whole 4$\pi$ solid angle. Pixel flux values are then summed together and divided by 4$\pi$ to obtain $J_{dust}$. 

\begin{figure}
 \vspace{-7.5mm}
 \hspace{-2mm}
  \includegraphics*[scale=0.5]{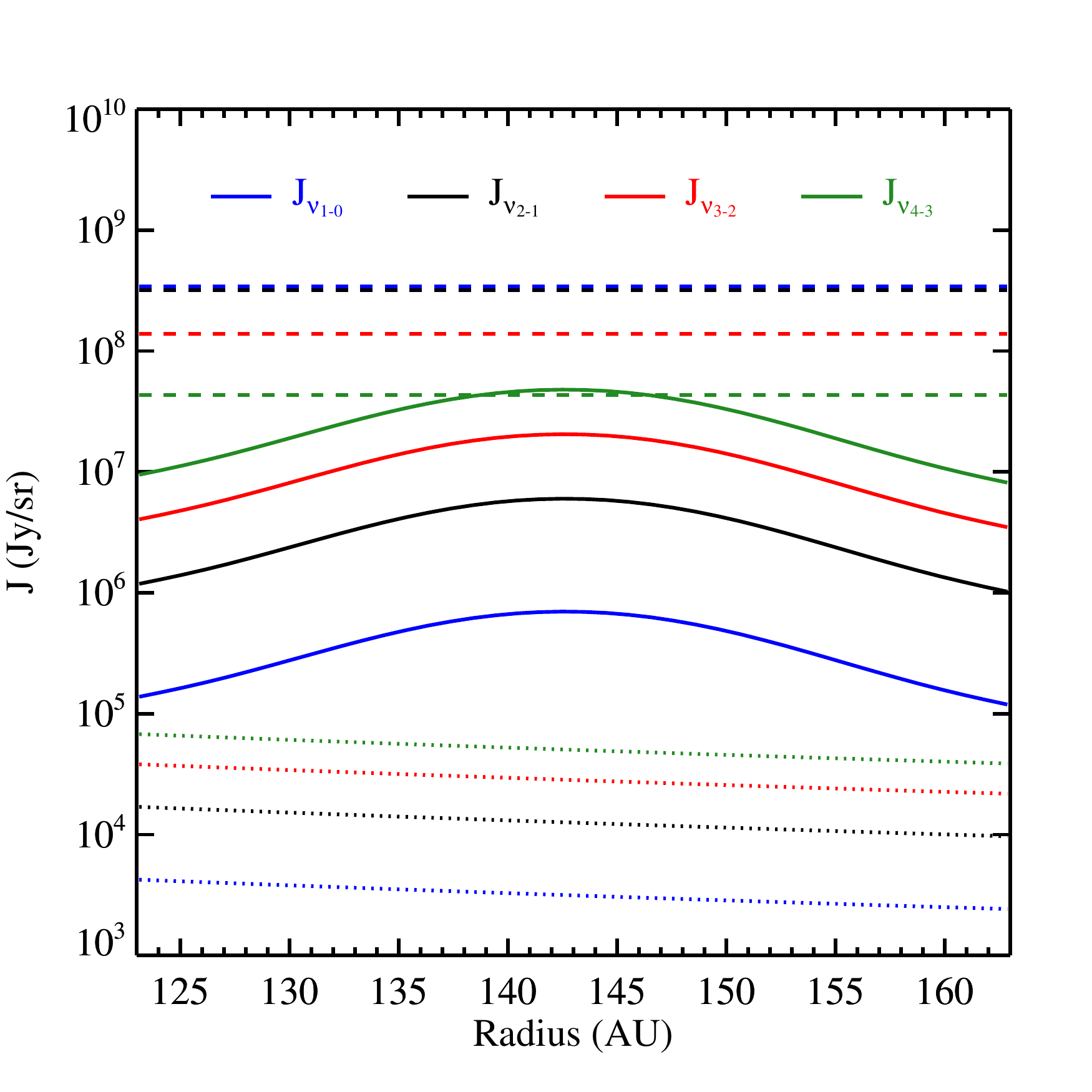}
\vspace{-7.5 mm}
  \caption{Components of the mean radiation field $J$ within the midplane of the Fomalhaut ring as a function of distance from the star, for the 1-0 (blue), 2-1 (black), 3-2 (red) and 4-3 (green) rotational transitions. Solid, dashed, and dotted lines represent the dust, CMB and stellar contributions, respectively.}
\label{fig:3}
\end{figure}

The different contributions to the radiation field within the Fomalhaut ring (at the frequency of the lowest 4 rotational transitions) are shown in Fig. \ref{fig:3}.
We see that the dust contribution is small compared to the CMB for the 1-0, 2-1 and 3-2 transitions. For the 4-3 and higher transitions the dust component becomes increasingly dominant, also due to the decreasing strength of the CMB field towards higher frequencies. The stellar contribution $J_{star}$ remains negligible compared to $J_{dust}$ and $J_{cmb}$ at all wavelengths. Finally, we note that the shape of $J_{dust}$ is slightly skewed towards shorter radii; this is attributable to radiation originating from dust on the opposite side of the ring with respect to the star.

\subsection{Collisional excitation}
\label{Sect:collexc}

As previously mentioned, the density of collisional partners $n$ is assumed to have the same radial and vertical distribution as the CO gas and the dust (where the latter is obtained from best-fit to the sub-mm continuum). We can therefore choose a total mass of collisional partners $M_{cp}$ and obtain $n$ at each point in the disc, where this, in turn, will influence the local excitation of the CO molecule. To ease interpretation, we leave the collisional partner-to-dust mass ratio $f_{cp}$ = $M_{cp}/M_{dust}$ \citep[where $M_{dust}=0.017$ M$_{\oplus}$ is the total dust mass in the disc,][]{Boley2012} as the free parameter, rather than $M_{cp}$ or $n$.

Unfortunately, collisional rate coefficients $\gamma$ are not known for collisions between CO and any gaseous species other than H$_2$, H$_2$O and electrons. This is because cases where these species are not the main collisional partners are rare in well-studied environments such as, for example, the interstellar medium (ISM) and protoplanetary discs. In debris discs, however, we lack observational constraints on the composition of the gas, meaning that there could be other species playing a non-negligible role in the collisional excitation of CO. We here model two different hypothetical environments, which drive the choice of collisional partner.

In the first scenario, the gas is assumed to retain its primordial composition and hence to contain a significant amount of H$_2$, which acts as the dominant collisional partner. Coefficients for H$_2$-CO collisions are obtained from the LAMDA database \citep{Schoier2005}.

Conversely, in the second hypothetical scenario, we assume the gas to originate from ices in planetesimals, and therefore to contain a mixture dominated by CO, H$_2$O, and their main photodissociation products. In fact, once released, each CO molecule photodissociates into C and O, where carbon will be subsequently ionised to C$^+$ in very short timescales, while O will remain mostly neutral due to its high ionisation potential \citep[e.g.,][]{Zagorovsky2010}. H$_2$O, on the other hand, will photodissociate mainly into H and OH, with OH in turn yielding atomic species O and H. Similarly to oxygen, atomic hydrogen is also expected to remain in the neutral phase, since the flux of photons from the interstellar UV field \citep[e.g.][]{Draine1978} at wavelengths below its ionisation threshold (912 \AA) is insufficient to keep H ionised. The same applies for the stellar flux from an A star such as Fomalhaut, since this is steeply decreasing at such short wavelengths (e.g., at 912 \AA$ $ the flux is already $>8$ orders of magnitude lower than its peak value).
Here we treat the electrons released from C ionisation \citep[as suggested by][]{Zuckerman2012} and the H$_2$O released by planetesimals as the dominant collisional partners; other potential colliders (such as atomic photodissociation products and CO itself) would serve to increase the collisional rates. 

In this scenario, the presence of two colliders introduces their relative abundance $n_{e^-}/n_{H_2O}$ as an extra free parameter in our calculations. Using reasonable assumptions, however, we can relate this ratio to the $n_{CO}/n_{H_2O}$ ice ratio in planetesimals. Assuming that CO and H$_2$O are released solely from icy planetesimals, then the steady-state gas-phase abundance ratio is related to the relative gas lifetimes and the ice abundances:
\begin{equation}
\left(\frac{n_{CO}}{n_{H_2O}}\right)_{gas}=\frac{\tau_{CO}}{\tau_{H_2O}}\left(\frac{n_{CO}}{n_{H_2O}}\right)_{ice},
\label{eq:lifetimes}
\end{equation}
where $\tau_{CO}$ and $\tau_{H_2O}$ are, in our case, the photodissociation lifetimes of CO ($\sim$120 yrs, see Sect. \ref{Sect:sign}) and H$_2$O gas ($\sim$6.2 $\times$ 10$^{-3}$ yrs, see Sect. \ref{Sect:sign}), respectively. Considering the lack of observational constraints in the Fomalhaut system, we assume C/CO and C$^+$/C ratios similar to those inferred for the $\beta$ Pictoris disc \citep[i.e. $\sim$100 and $\sim$1 respectively,][]{Roberge2000, Cataldi2014}. Assuming the electron abundance is dominated by the photodissociation of C, then:
\begin{equation}
\frac{n_{e^-}}{n_{H_2O}}=\frac{n_{C^+}}{n_{C}}\frac{n_{C}}{n_{CO}}\left(\frac{n_{CO}}{n_{H_2O}}\right)_{gas}\approx 100\left(\frac{n_{CO}}{n_{H_2O}}\right)_{gas} .
\label{eq:ratios}
\end{equation}
Together with Eq. \ref{eq:lifetimes}, Eq. \ref{eq:ratios} allows us to leave the $(n_{CO}/n_{H_2O})_{ice}$ ratio in planetesimals as the extra free parameter. This is extremely useful, as it enables us to interpret our results self-consistently in terms of the volatile component of planetesimals in the Fomalhaut system (see Sect. \ref{Sect:ressec}). Collisional rate coefficients for $\mathrm{e^-}$-CO collisions are obtained using expressions described by \citet{Dickinson1975}, whereas collisional rate coefficients for H$_2$O-CO collisions are calculated following the method outlined in \citet{Green1993}.

As well as the collisional partner-to-dust mass ratio $f_{cp}$ (and, for the secondary scenario, the CO/H$_2$O ice ratio), the kinetic temperature of the gas ($T_k$) is also left as a free parameter. Its local value depends on the heating/cooling balance and on the chemistry within the ring. Detailed chemical modelling is beyond the scope of this paper; due to the relatively small spatial extent of the ring, we assume $T_k$ to be uniformly distributed, and to be equal for all colliders. Again, in order to probe a wide region of parameter space, we select a range of temperatures from as low as 10 K, to as high as 150 K ($\sim$3 times the dust temperature inferred by \citet{Boley2012}).

In summary, $f_{cp}$, $T_k$ and $(n_{CO}/n_{H_2O})_{ice}$ are left as free parameters, and the set of $j_{max}$+1 statistical equilibrium equations (Eq.~11 and ~12) is solved to obtain fractional populations $x_j$ for all levels up to $j_{max}$=24. The upper limit on the integrated line flux is then used to derive an upper limit on the total CO mass using Eq.~2. These results must then be compared with any compositional assumptions; e.g., implicit in the secondary gas model is a prediction for the mass of CO present for a given $f_{cp}$ and $(n_{CO}/n_{H_2O})_{ice}$, which can thus be compared with the derived upper limit on the CO mass.

\subsection{Results for the primordial gas scenario}
\label{Sect:resfir}
 
\begin{figure}
 \vspace{-5.5mm}
 \hspace{-5mm}
  \includegraphics*[scale=0.46]{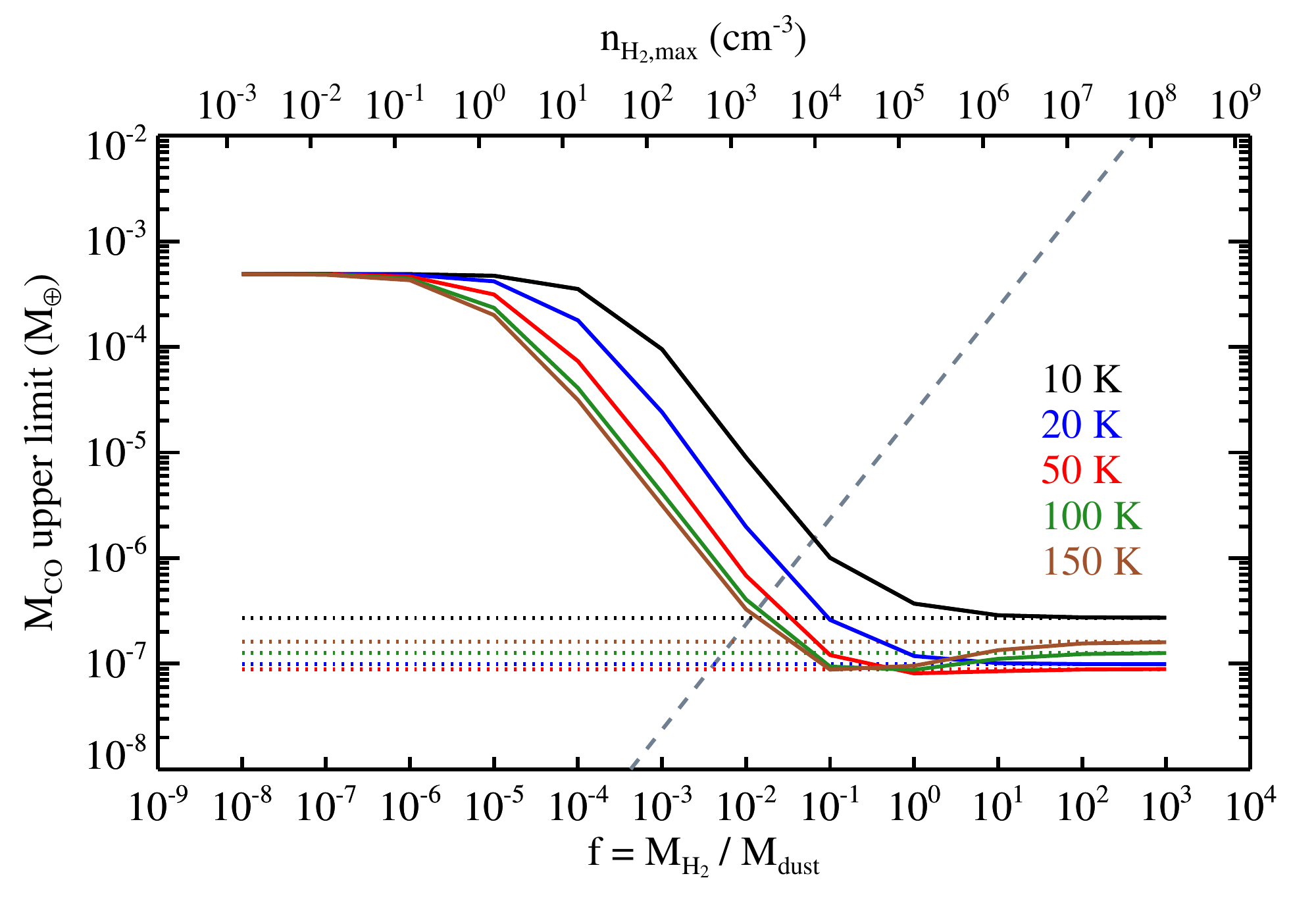}
\vspace{-7 mm}
  \caption{Primordial scenario, with H$_2$ as the dominant collisional partner. Upper limits on the CO mass present in the Fomalhaut ring, as a function of the collisional partner-to-dust mass ratio $f_{cp}$ and gas kinetic temperature $T_k$ (different colours). Solid lines show results in the most general scenario (from Eq.~4), whereas dotted lines show results in the LTE regime (from Eq.~5). The grey dashed line represents a constant CO/H$_2$ gas abundance ratio of 10$^{-4}$, typical of the ISM. }
\label{fig:4}
\end{figure}
For the primordial scenario (i.e. collisions with H$_2$), results of our calculations are shown in Fig.\ref{fig:4}. It is apparent that LTE (dotted lines) applies when the density of collisional partners is high compared to the critical density (see Eq.~8). As this regime is approached, we note in Fig. \ref{fig:4} an inversion in the behaviour of the level populations, and hence of our mass upper limits, at different kinetic temperatures. This is because in LTE the $j$=3 level is populated most at the temperature corresponding to its characteristic energy (from $T$ = $E_{j=3}/k$ $\simeq$ 33 K), and hence it is for this temperature that the least CO is needed to produce the observed flux upper limit. At very low collisional partner densities, on the other hand, we are in the radiation-dominated regime (see Eq.~10): the CO mass estimate loses its dependence on both the temperature and the local density of the colliders, depending only on the continuum radiation field, which in the case of Fomalhaut is accurately known.

At both extremes, the results are independent of the density of collisional partners (as expected from Eq. 8 and 10) allowing, in case of a detection, to set strict, collider-independent upper and lower limits on the CO content in the disc. In the environment of the Fomalhaut ring, and for the transition 3-2, it is in the radiation-dominated regime that most mass is needed to produce the observed flux; the 3$\sigma$ upper limit on the mass of CO in the Fomalhaut ring is thus 4.9 $\times$ $10^{-4}$ M$_{\oplus}$. Interestingly, for Fomalhaut this result depends only on the strength of the CMB field at the frequency of the transition (see Sect. \ref{sect:radenv}).

Here we consider whether there are any additional considerations which allow us to constrain the parameter space of the model shown in Fig.\ref{fig:4}.
In our models, LTE is reached as the collisional partner gas/dust ratio $f_{cp}$ approaches 100, consistent with the value found in the interstellar medium (ISM). Although a gas/dust ratio $f_{cp}$ of $100$ (and LTE) is the generally assumed value for protoplanetary discs, the inferred upper limit on the CO/H$_2$ abundance at $f_{cp}=100$ is more than 3 orders of magnitude lower than the ISM value of 10$^{-4}$, which is also typically assumed to be the case in protoplanetary discs. On the other hand, an ISM CO/H$_2$ abundance (see dashed line on Fig. \ref{fig:4}) is possible, but only if $f_{cp}<0.1$ (in the highly non-LTE regime). While the above arguments require any primordial gas to have either a non-primordial abundance ratio or a non-primordial gas/dust ratio, there is no apparent reason for these ratios to be retained at an age as advanced as that of Fomalhaut ($\sim$440 Myr). Thus gas that is primordial in origin could still be present as long as some process has acted to reduce its CO/H$_2$ ratio or to reduce the overall gas mass relative to the dust (although Section \ref{Sect:sign} gives other reasons why retention of primordial gas is unlikely).

\subsection{Results for the secondary gas scenario}
\label{Sect:ressec}

\begin{figure*}
\vspace{-10mm}
\centering
 \hspace{-20mm}
 \begin{subfigure}{0.45\textwidth}
  \includegraphics*[scale=0.47]{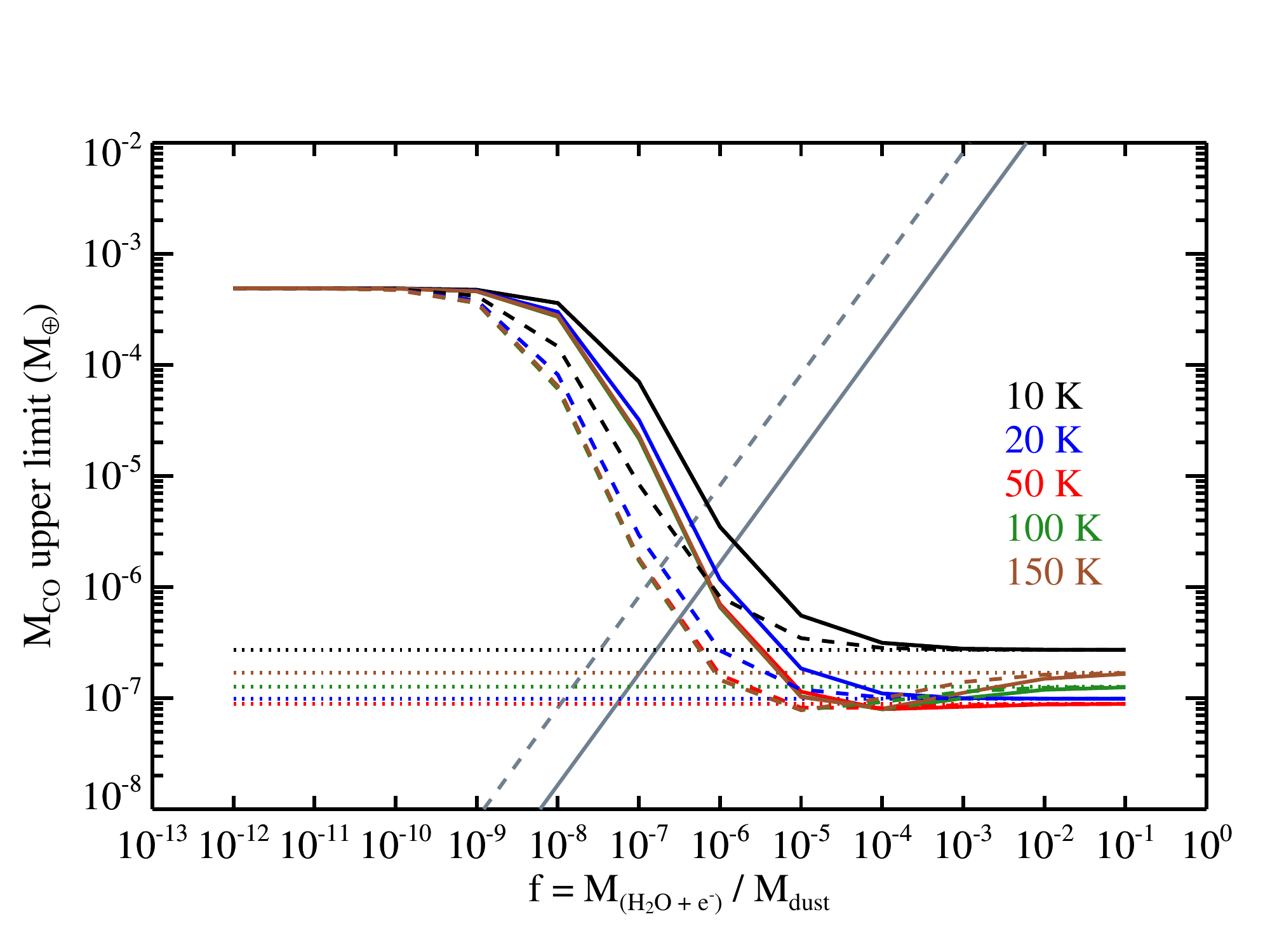}
  \end{subfigure}
  \hspace{12mm}
  \begin{subfigure}{0.45\textwidth}
   \includegraphics*[scale=0.47]{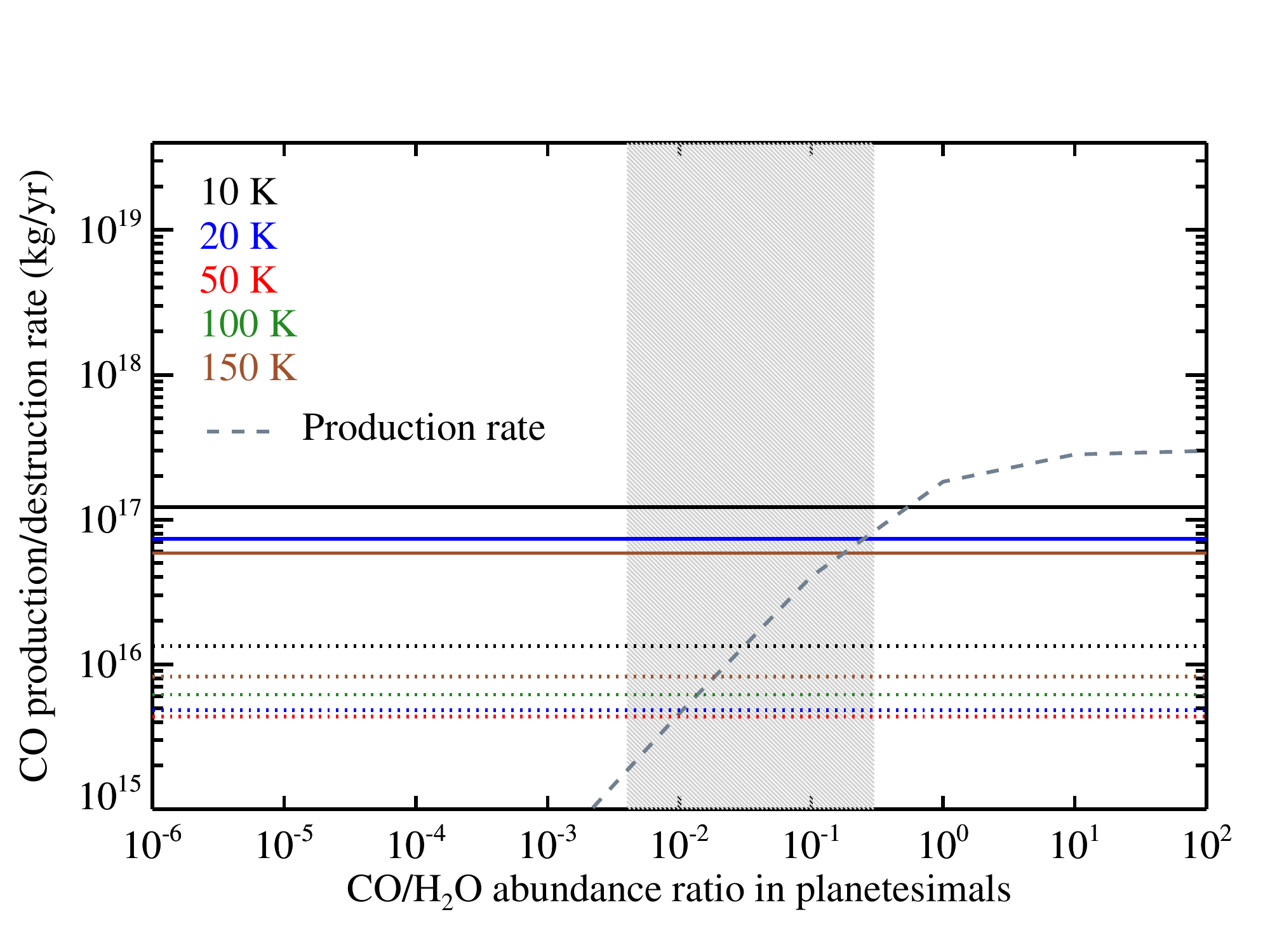}
   \end{subfigure}
   \vspace{-3mm} 
   \caption{Secondary scenario. Left: upper limits on the CO mass present in the Fomalhaut ring, as a function of the collisional partner-to-dust mass ratio $f_{cp}$ (x axis, where the main colliders are H$_2$O and $\mathrm{e^-}$), of the CO/H$_2$O ice abundance ratio in planetesimals (different line styles) and of the gas kinetic temperature $T_k$ (different colours). The CO/H$_2$O ice ratios used to obtain the displayed upper limits (see Sect. \ref{Sect:collexc} for details) are 0.4\% (solid curves) and 30\% (dashed curves). These are the boundaries of the range of values observed in Solar System comets \citep{Mumma2011}. Diagonal lines of constant CO/H$_2$O gas ratios are shown in grey, and were derived for the same limiting ice ratios through Eq. 13. Dotted lines are upper limits assuming LTE regime. Right: CO steady state production/destruction rates as a function of the CO/H$_2$O ice abundance ratio within planetesimals. Solid lines represent observational upper limits on the destruction rate, where these correspond to the CO mass upper limits at the intersections between grey and coloured lines on the left plot, divided by the CO photodissociation timescale of 120 years. Dotted lines represent upper limits on the destruction rate if LTE is assumed. The dashed grey curve shows the CO production rate derived from the rate at which mass is passed down the collisional cascade \citep[$6.0\times10^{17}$ kg/yr,][]{Wyatt2002}, for different ice compositions. The shaded area represents the whole range of abundance ratios observed in Solar System comets.}
\label{fig:5}
\end{figure*}

For the secondary generation scenario (i.e. collisions with electrons and H$_2$O), results of our calculations for the two extremes of the range of CO/H$_2$O ice ratios observed in Solar System comets \citep[solid lines: 0.4\% and dashed lines: 30\%,][]{Mumma2011} are shown in Fig. \ref{fig:5}, left. Implicit in the secondary model is that the mass of CO present in the gas disc is fixed for a given assumption about $f_{cp}$ and $(n_{CO}/n_{H_2O})_{ice}$ in planetesimals. This is shown as the diagonal grey lines on Fig. \ref{fig:5} left, calculated from Eq. 13 and 14. Therefore, while the coloured lines of CO mass upper limits on that figure are instructive, the only points that are internally consistent in the model are those where the CO mass upper limits cross the grey line for the corresponding $(n_{CO}/n_{H_2O})_{ice}$ in planetesimals. Those intersection points are shown on Fig. \ref{fig:5} right, where they are plotted as CO destruction rates assuming a photodissociation lifetime of $\sim$120 years (see discussion in Sect. \ref{Sect:sign}). This shows that CO/H$_2$O ice ratios observed in Solar System comets (0.4-30\%) lead to upper limits on the CO and H$_2$O gas mass in the Fomalhaut ring of 2.5 $\times$ $10^{-6}$ M$_{\oplus}$ and 1.21-0.08 $\times$ $10^{-6}$ M$_{\oplus}$, respectively. 

The upper limit on the total CO mass found here is $>$ 2 orders of magnitude lower than our most conservative estimate (4.9 $\times$ $10^{-4}$ M$_{\oplus}$, in the radiation-dominated regime). This is because in our secondary gas scenario the presence of CO requires the presence of other species (in our case, electrons and water), which can act as collisional partners and push the molecular excitation away from the radiation-dominated regime. Despite the simplicity of the model, this resulting CO mass upper limit would be expected to apply more generally, since if anything the model underestimates the number of collisional partners, and any additional collisional partners would act to push towards LTE and so reduce this upper limit. If enough colliders were present for LTE to apply, the CO mass upper limit would be even lower, in turn pushing down the upper limits on the CO destruction rate (dotted lines in Fig. \ref{fig:5} right) and on the CO/H$_2$O ice ratio in planetesimals (see Sect. \ref{sect:compos}).

We note that our upper limits displayed as coloured curves in Fig. \ref{fig:5} (left) show the same dependence of the CO mass upper limit on $f_{cp}$ as those in Fig. \ref{fig:4}; i.e. transitioning from a radiation-dominated regime at low $f_{cp}$ to LTE at large $f_{cp}$. In fact, the same is true for any choice of collisional partner, there is just a difference in the range of $f_{cp}$ at which this transition between LTE and radiation-dominated regimes occurs.
As explained in Sect. \ref{Sect:molexc}, the ratio $n_{crit_{u,l}}/n$ is what determines where this transition takes place, i.e. whether the CO excitation is dominated by radiation, collisions (LTE), or a mixture of both. If, as in our case, two collisional partners are present, we have
\begin{equation}
\frac{n_{crit_{u,l}}}{n}=\frac{A_{u,l}}{n_{e^-}\gamma_{e^-_{u,l}}+n_{H_2O}\gamma_{H_2O_{u,l}}},
\label{eq:ncrovern}
\end{equation}
where $A_{u,l}$, $\gamma_{e^-_{u,l}}$ and $\gamma_{H_2O_{u,l}}$ are fixed, leaving $n_{crit_{u,l}}/n$ dependent on $n_{e^-}/n_{H_2O}$ and hence on $(n_{CO}/n_{H_2O})_{ice}$ only. Two limiting regimes are then present, where the collisional excitation of CO is dominated by either of the two species. Electrons will dominate in the limit where Eq. 15 becomes 
\begin{equation}
\frac{n_{crit_{u,l}}}{n}\sim\frac{A_{u,l}}{n_{e^-}\gamma_{e^-_{u,l}}}=\frac{n_{crit_{u,l}, e^-}}{n_{e^-}}.
\label{eq:ncrovern}
\end{equation}
Since we have $\gamma_{e^-_{u,l}}/\gamma_{H_2O_{u,l}}\sim10^2$, this limit is reached for $n_{e^-}/n_{H_2O}\gg10^{-2}$, or equivalently (using Eq. 13 and 14), for $(n_{CO}/n_{H_2O})_{ice}\gg5\times10^{-9}$. Conversely, H$_2$O will be the dominant collider in the limit $(n_{CO}/n_{H_2O})_{ice}\ll5\times10^{-9}$. This means that the CO/H$_2$O ice ratios probed in Fig. \ref{fig:5} left  (0.004 and 0.3) are well into the regime where electrons dominate collisional excitation of CO. Hence, changing the CO/H$_2$O ice ratio does not affect our resultant CO mass upper limit (again, found at the intersections between grey and coloured lines in Fig. \ref{fig:5} left). In turn, this also explains why the CO destruction rate upper limits (Fig. \ref{fig:5} right) are independent of the CO/H$_2$O ice ratio in planetesimals.

\section{Discussion}

Detailed analysis of the excitation of the rotational levels in CO molecules has enabled us to convert our ALMA integrated flux upper limit to upper limits on the CO gas mass and on the CO/H$_2$O ice ratio within planetesimals in the Fomalhaut ring. In addition, it underlined the importance and influence of the gaseous environment on CO excitation and hence line emission in a debris disc. In the following sections, we will discuss the consequence of our analysis on the presence of CO in Fomalhaut and other debris systems. Finally, we will consider prospects for its future detection and characterisation using ALMA.

\subsection{Production and destruction of gas in the Fomalhaut ring}
\label{Sect:sign}

In the scenario where gas is of second generation, our results have shown that less than 2.5 $\times$ 10$^{-6}$ M$_{\oplus}$ of carbon monoxide are present in the Fomalhaut ring. In this section, we aim to understand if such low amounts of CO in the Fomalhaut ring can be explained by any of the processes that have been so far deemed responsible for destruction and production of CO gas in debris discs. 

Depletion of gas can take place via both photodissociation and freeze-out onto grains. Photodissociation is triggered by discrete absorption of UV photons originating from both the star and the interstellar radiation field (ISRF), in the wavelength range 911.75-1117.80 \AA$ $ \citep{Visser2009}.
We ignore the effect of dust, H$_2$ and self- shielding, since these are expected to be negligible at the low dust and gas densities within the Fomalhaut ring \citep{Kamp2000}. For the ISRF UV contribution in the wavelength range described above, we adopt the \citet{Draine1978} field as described by Eq. 3 in \citet{vanDishoeck1988}. 
For the stellar UV field, we used a best-fit stellar spectrum model of Fomalhaut (i.e. with stellar surface gravity log $g=3.52$, effective temperature T$_{eff}=8560$ K, and metallicity [M/H]=0.22). We then used cross sections from \citet{Visser2009} to calculate the photodissociation rate; we find that the stellar contribution is negligible at the distance of the Fomalhaut ring ($\sim$143 AU) and hence assume the photodissociation timescale of CO to be $\tau_{phd}\sim120$ years \citep{Visser2009}, typical of the ISRF. Such short timescale makes a primordial origin for CO extremely unlikely, unless a suitable mechanism can be found to shield CO from ambient radiation for 440 Myr. As such, primordial origin gas is not favoured in this system (see also Sect. \ref{Sect:resfir}).

Compared to CO, the UV range over which H$_2$O photodissociation can be triggered extends to longer wavelengths (up to 2000 \AA), and the cross sections are generally higher \citep{Lee1984}. This means that the water molecule is much easier to photodissociate, particularly since the stellar flux increases very steeply at these longer wavelengths. Indeed, we find that the stellar flux dominates photodissociation; the timescale at $\sim$143 AU is 6.2 $\times$ 10$^{-3}$ years (2.3 days).

On the other hand, the timescale for freeze-out of molecules onto grains can be expressed by \citep[e.g.,][]{Flower2005}
\begin{equation}
\tau_{fo}=[S(T_{gr})\sigma_{gr}v_{th}]^{-1}
\end{equation}
where $v_{th}$ is the thermal velocity of the molecules, given by 
\begin{equation}
v_{th}=\left(\frac{8k_BT_{k}}{\pi m_{CO}}\right)^{\frac{1}{2}},
\end{equation}
$S(T_{gr})$ is the sticking coefficient, and $\sigma_{gr}=n_{gr}\pi r_{gr}^2$ is taken as the total cross section per unit volume of the grains, whose value varies within the ring. In order to find the shortest possible $\tau_{fo}$, we choose both the highest value of $\sigma_{gr}$ from our dust model and the highest $T_{k}$ in the range explored in Sect. 5. Given a dust temperature of 48 K \citep{Boley2012}, we adopt a sticking coefficient of $\sim$0.4 \citep[See Fig. 1,][]{Sandford1990}. The shortest timescale for CO freeze-out within the ring is $\tau_{fo}\sim6\times10^4$ years, much longer than $\tau_{phd}\sim120$ years, meaning that photodissociation dominates the depletion of CO gas. A similar argument applies for H$_2$O gas, as the only difference in the rate with respect to CO is the dependence on the molecular weight.

We now turn to look at processes that could produce CO and H$_2$O within the ring. In particular, we explore the possibility of a cometary origin, such as inferred in the $\beta$ Pic system \citep[e.g.,][]{Dent2014}. Release of these gases from comets is attainable via processes such as sublimation, collisions and photodesorption. At a dust temperature of $\sim$50 K, the timescale for thermal desorption (sublimation) of CO deposited on top of H$_2$O ice is estimated to be $2\times10^{-5}$ yr \citep[i.e. $\sim11$ minutes,][]{Sandford1990}, meaning that all the CO residing on the surface of comets in the Fomalhaut ring was lost early in the lifetime of the ring. On the other hand, at this temperature, H$_2$O will remain in the solid phase, allowing some CO to remain trapped within it. \citet{Collings2003} carried out laboratory experiments looking at the desorption of CO that was at first deposited either on top of, or below H$_2$O ice. According to their study, up to $\sim$50\% of the CO that originally froze onto grains can still be trapped within the water ice layer of comets in the Fomalhaut ring at $T_{dust}=48$ K. However, the study also shows that the degree of entrapment also depends on the temperature at which H$_2$O froze itself, so it is important to keep in mind that this will depend on where in the protoplanetary disc water formed and froze onto grains. We therefore assume that a reservoir of CO can be present at late ages in a planetary system even within its ice line, trapped within H$_2$O in cometary bodies.

We now analyse whether release of this entrapped CO is possible, and quantify it using simple assumptions. In the Fomalhaut ring, the rate at which large planetesimals in the belt are being ground down, which is the same as the rate at which small dust grains are being replenished, has been estimated as $6.0\times10^{17}$ kg/yr \citep{Wyatt2002}. Assuming the planetesimal composition to be similar to that observed in Solar System comets (ice/rock ratio of about 1, CO/H$_2$O ice ratios between 0.4 and 30\%), we can use it to estimate the rate of CO release, which corresponds to between $1.9\times10^{15}$ and $8.0\times10^{16}$ kg/yr. The mechanisms that can be responsible for this release are collisions and photodesorption. Collisions can contribute through dust vaporisation \citep{Czechowski2007}, planetesimal break-up \citep{Zuckerman2012} and giant impacts \citep{Jackson2014}. Photodesorption, on the other hand, will affect the H$_2$O ice on the surface of solids, in turn exposing CO and allowing it to escape on very short timescales. The rate at which water vapour will be released is $da/dt\sim1.5\times10^{-3}$ $\mu$m/yr \citep[scaled to the distance of the ring, from the result of][]{Grigorieva2007}, where $a$ is the vertical thickness of the layer. The total water mass released in this manner will then be
\begin{equation}
\frac{dM_{H_2O}}{dt}=4\sigma_{tot}\rho_{gr} \frac{da}{dt}
\end{equation}
where $\rho_{gr}$ is the grain density, which we take as that of water ice ($\sim1\times10^{-15}$ kg $\mu$m$^{-3}$), and $\sigma_{tot}$ is the total cross section of icy grains in the Fomalhaut ring (in $\mu$m$^2$). Under the assumption that all the grains are fully icy, we can use the total cross sectional area of the Fomalhaut ring \citep[33.7 AU$^2$,][]{Wyatt2002} to obtain an H$_2$O production rate of $4.6\times10^{18}$ kg/yr. This is higher than the rate at which mass is being passed down the collisional cascade (again, $6.0\times10^{17}$ kg/yr), pointing towards a more realistic scenario where grains are not fully icy, but made up of a mixture of ice and rock. We therefore conclude that photodesorption of H$_2$O might contribute significantly to the release of trapped CO gas from planetesimals in the ring, though the extent of this contribution depends on how much of the cross-sectional area of the Fomalhaut disc is icy. In any case, the exact mechanism for CO gas production is unimportant as long as there is one that can feasibly explain its release. 

\subsection{Constraints on the planetesimal composition in the secondary gas model}
\label{sect:compos}

In the previous section, we estimated CO production rates as a function of CO/H$_2$O ice ratio from the rate at which planetesimals are being ground down through the collisional cascade (i.e. CO production rate = $6.0\times10^{17}$ $\times$ $(n_{CO}/n_{H_2O})_{ice}$ kg/yr). In Sect. \ref{Sect:ressec} we also calculated upper limits on the destruction rates, by combining our mass upper limits from ALMA observations to the photodissociation timescale within the ring. Unless any secondary CO were produced by a transient event, it would have to be replenished at a steady rate. If we assume steady state, the production rate (grey dashed line, Fig. \ref{fig:5}, right) should balance the destruction rate (coloured lines), allowing us to set an upper limit of $\sim$55\% on the CO/H$_2$O ice ratio in planetesimals within the Fomalhaut ring. It therefore remains plausible for CO gas to be actively released by Solar-System-like cometary bodies within the ring, but yet not to have been detected so far.
If however LTE holds, which would be true if our model underestimated the number of colliders present, the derived CO destruction rate would be much lower (see dotted lines in Fig. \ref{fig:5} right and Sect. \ref{Sect:ressec}). This would set a much stronger constraint on the planetesimal ice composition (CO/H$_2$O $\lesssim3$\%), pushing towards the low end of the range observed in Solar System comets (0.4-30\%).

\subsection{Comparison with the $\beta$ Pictoris disc}

If we assume that any gas present in the Fomalhaut ring is of secondary origin, it is instructive to draw a comparison with $\beta$ Pictoris, where secondary gas has been detected and resolved \citep{Dent2014}. The two systems contain similar amounts of sub-millimetre dust mass \citep[0.017 M$_{\oplus}$ in the Fomalhaut ring as compared to 0.08 M$_{\oplus}$ in the $\beta$ Pic disc; e.g.][]{Boley2012,Dent2014}, though our upper limit CO flux, once scaled to the same distance as $\beta$ Pic, is lower by about three orders of magnitude (see Sect. \ref{sect:prosother}). We here suggest two possible explanations.

The first is a higher rate of collisions within the $\beta$ Pic disc, consequently releasing higher amounts of CO. The published collision rates are similar: $6.0\times10^{17}$ kg/yr in the Fomalhaut case as compared to $6.3\times10^{17}$ kg/yr for $\beta$ Pic \citep{Czechowski2007}; however, the latter does not take into account the presence of a clump in the $\beta$ Pic disc \citep[as observed by][]{Dent2014}. If this is due to resonances, then the dynamics results in a greatly enhanced collision rate \citep[][]{Wyatt2006}. The same also applies if the clump originates from a recent giant impact \citep{Jackson2014}. We would therefore expect the collision rate within the $\beta$ Pic disc to be significantly higher than in the Fomalhaut ring, thus releasing significantly more gas.

Another explanation might be the age of the Fomalhaut system ($\sim$440 Myr), much more advanced than the $\beta$ Pic system \citep[$\sim$21 Myr,][]{Binks2014}. In fact, collisions and photodesorption can strip ices off the surface of dust grains and planetesimals on timescales that are short \citep[and increasingly shorter for smaller solids, see e.g.][]{Grigorieva2007} compared to the age of the Fomalhaut system, so it is not unreasonable for solid material in a debris disc to have lost most of its ice by 440 Myr, while still retaining a considerable amount at 21 Myr. For example, in the Fomalhaut system planetesimals may have lost most of their ice mantles by 440Myr due to photodesorption, leaving the collisional grinding as the only possible channel to release any further ice that may be trapped inside them.

However, we stress that care must be taken when associating CO fluxes with masses (see Sect. 5.5). If we consider that $\beta$ Pic has a total dust mass similar to that found in the Fomalhaut ring, but spread in a much broader disc, we anticipate its dust radiation field within the disc to be lower than found in Fomalhaut, and the total radiation field at the frequency of the lowest rotational transitions of CO to be still dominated by the CMB. This means that the CO $j=3$ level will be populated in exactly the same way for both systems, implying a similar discrepancy in the derived CO masses between the LTE and radiation-dominated regimes of $\sim$3-4 orders of magnitude (depending on the gas temperature). To first approximation, this means that if the collisional partner density in the $\beta$ Pic disc is low enough, up to $\sim$0.03-0.3 M$_{\oplus}$ of CO could be present in its disc, as compared to the $\sim$$3\times10^{-5}$ M$_{\oplus}$ derived for the LTE regime \citep{Dent2014}. It is also plausible, however, that the collisional partner density in $\beta$ Pic is high enough for LTE to be a good approximation, considering that several other atomic and ionic gaseous species have already been detected in the system \citep[e.g.,][]{Roberge2000, Cataldi2014}. If this were to be the case, and if on the other hand the CO in Fomalhaut were in the radiation-dominated excitation regime, it would still be possible for the two systems to contain similar amounts of CO mass, despite the considerable discrepancy in their line flux. In turn, however, this would raise the question of why several species other than CO are present in $\beta$ Pic, but not in Fomalhaut.

\subsection{Prospects for detection with ALMA in the Fomalhaut ring}
\begin{figure*}
\vspace{-5mm}
\centering
 \hspace{-12mm}
 \begin{subfigure}{0.45\textwidth}
  \includegraphics[scale=0.47]{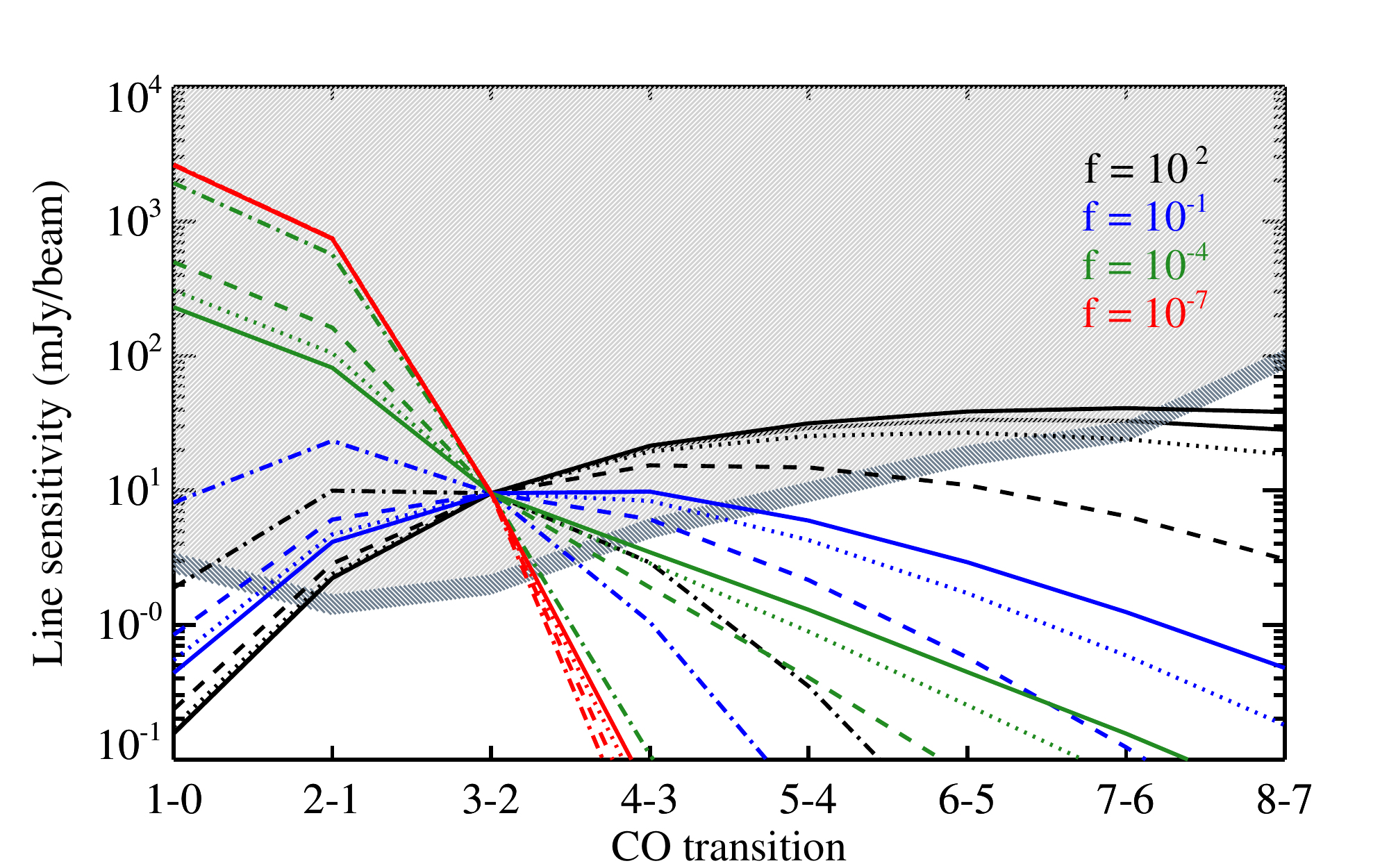}
  \end{subfigure}
  \hspace{0mm}
  \begin{subfigure}{0.45\textwidth}
   \includegraphics[scale=0.47]{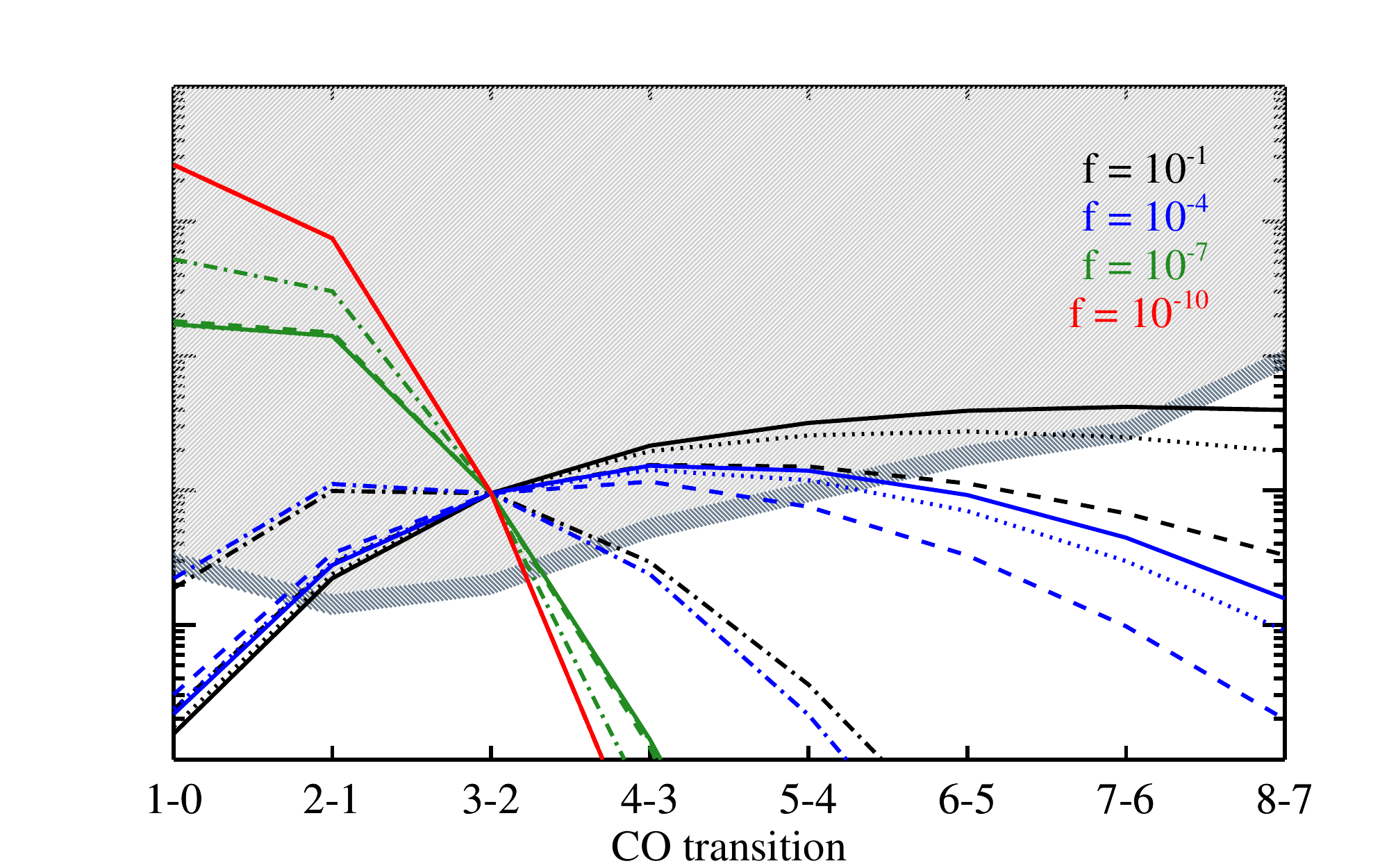}
   \end{subfigure}
   
  \caption{Sensitivity per beam per spectral channel (1$\sigma$) required to probe the same amount of mass in the Fomalhaut system as our upper limit (i.e. 4.9 $\times$ $10^{-4}$ M$_{\oplus}$ for small $f_{cp}$, and $\sim$ $10^{-7}$ M$_{\oplus}$ for large $f_{cp}$), for different CO rotational transitions. Different colours represent different collisional partner to dust mass ratios $f_{cp}$, while different line styles represent different gas kinetic temperatures $T_k$ (solid: 100 K, dotted: 70 K, dashed: 40 K, dash-dotted: 10 K). Left: collisions with H$_2$. Right: collisions with H$_2$O and electrons, for a cometary CO/H$_2$O abundance ratio of 0.4\%. We maximise detectability by choosing a circular beam of FWHM equal to approximately the width of the continuum ring ($\sim$2$\arcsec$) for transitions 1-0 to 3-2, whereas for higher transitions we choose resolution ($<$2$\arcsec$) corresponding to the most compact ALMA configuration. 
The velocity resolution is the same as in the observations treated in this paper (0.424 km/s). Shaded areas represent emission levels detectable in the future using ALMA (light grey: 1 hour on-source, dark grey: 2 hours).}
\label{fig:6}
\end{figure*}

We now examine the prospect of future detectability of CO in the Fomalhaut ring, using the capabilities of the full ALMA array. For transitions 1-0 up to 3-2, we select a beam of FWHM equal to the width of the ring ($\sim$2$\arcsec$) in order to achieve maximum detectability. For higher transitions, beam sizes obtainable by ALMA in its most compact configuration are chosen (4-3:1.8$\arcsec$, 6-5:1.2$\arcsec$, 7-6:1.0$\arcsec$, 8-7:0.9$\arcsec$). We note that the 5-4 transition is not observable with ALMA, because of terrestrial atmospheric absorption. For simplicity, we select a velocity resolution equal to the one in the observations treated in this paper (i.e. 0.424 km/s, see Sect. 3), though keeping in mind that the optimal resolution would correspond to the spread in ring velocities expected within a beam area, which will depend on the CO line frequency, and on the location within the ring. The sensitivity calculations are carried out using fifty 12-m array antennas, assuming precipitable water vapour values appropriate for each ALMA band, for integration times of 1 and 2 hours. Then, using Eq. 2, mass upper limits $M$($f_{cp}$, $T_k$) from Fig. \ref{fig:4} and \ref{fig:5} (left) are turned into integrated flux upper limits $F_{j,j-1}$($f_{cp}$, $T_k$) for the different CO rotational lines. Flux upper limits per spectral channel per beam are then derived using the same spatial and spectral resolutions as for the sensitivity calculations. 
Fig. \ref{fig:6} shows a comparison between these flux upper limits and the output of the ALMA Sensitivity Calculator.

In agreement with our assumptions, all lines intersect at the 3-2 transition: the value of sensitivity per spectral channel per beam at this frequency corresponds to that achieved in the data analysed in this paper (i.e. $\sim$3.4 mJy/beam per spectral channel), corrected for the difference in beam size and for the increased sky area occupied by the full ring as compared to the partial ring analysed in this work (see Fig. \ref{fig:1}).

It is evident that the widest range of plausible scenarios is probed for the lowest transitions (particularly for the 1-0 and 2-1). Therefore observations at low frequencies would be the most favourable for detection, particularly for low density of collisional partners, and low gas kinetic temperatures. This is to be expected if we consider that radiation alone is not sufficient to significantly excite CO molecules to high-$j$ levels in the Fomalhaut ring, while collisional excitation is able to do so. In addition, the sensitivity achievable with ALMA in a given observational time is substantially better at low frequencies, further increasing the likelihood of detection. Indeed, observing the CO 1-0 or 2-1 transitions in the Fomalhaut ring using ALMA (for 1-2 hours of integration time) could potentially detect CO masses down to the $\sim$10$^{-7}$ M$_{\oplus}$ level (depending on $n$ and $T_k$), corresponding to the total CO mass present in $\sim$10$^4$ comets Halley\footnote{Comet mass = 2.2$\times$10$^{14}$ kg \citep[][]{Hughes1985}, CO abundance $\sim$17\% \citep[][]{Woods1986}}.

\subsection{Prospects for detection with ALMA in other systems: is gas common in (young) debris discs?}
\label{sect:prosother}

\begin{figure}
 \vspace{-10mm}
 \hspace{-3mm}
  \includegraphics*[scale=0.52]{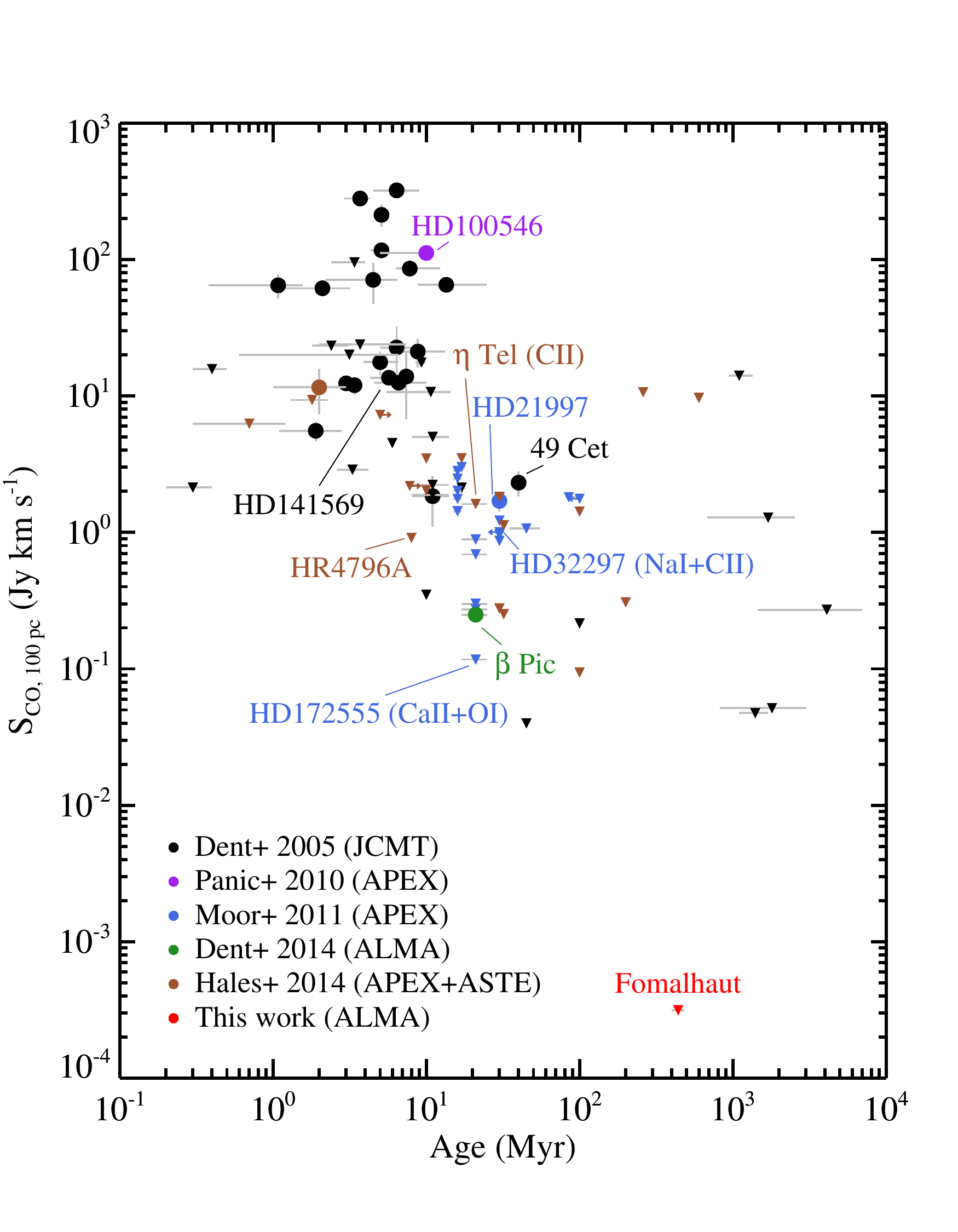}
\vspace{-10mm}
  \caption{CO 3-2 integrated line flux (normalised to a distance of 100 pc) against age for a sample of Herbig Ae/Be and debris discs. Filled dots are detections, inverted triangles are upper limits (1$\sigma$, for a line width of 10 km s$^{-1}$). The sample comprises sources observed in the CO surveys by \citet{Dent2005} (black symbols), \citet{Moor2011} (blue symbols) and \citet{Hales2014} (brown symbols), with the addition of extra sources observed in CO 3-2, namely HD100546 \citep[purple dot,][]{Panic2010}, $\beta$ Pic \citep[green dot,][]{Dent2014} and Fomalhaut (red triangle, this work). Fluxes, where retrieved in K km s$^{-1}$, have been converted to Jy km s$^{-1}$ using appropriate conversion factors, as stated within the references or on the relevant instrument webpages. Ages have been obtained from the following references: \citet{Zuckerman1995}, \citet{VandenAncker1998}, \citet{Mamajek2002}, \citet{Chen2005}, \citet{Hernandez2005}, \citet{Rieke2005}, \citet{Guimaraes2006}, \citet{Manoj2006}, \citet{Moor2006}, \citet{Rhee2007a}, \citet{Rhee2007b}, \citet{Hillenbrand2008}, \citet{Smith2008}, \citet{Torres2008}, \citet{Verrier2008}, \citet{AlonsoAlbi2009}, \citet{Montesinos2009}, \citet{Casagrande2011}, \citet{Vacca2011}, \citet{Folsom2012}, \citet{Lisse2012}, \citet{Mamajek2012}, \citet{Nakajima2012}, \citet{Pecaut2012}, \citet{Zuckerman2012}, \citet{Alecian2013}, \citet{Binks2014}.}
\label{fig:7}
\end{figure}
Not only can ALMA be used to improve our detection limit for Fomalhaut, but also to probe down to much deeper CO emission levels for other discs. In Fig.\ref{fig:7}, we show integrated CO J=3-2 intensity against age for a number of debris and Herbig Ae/Be discs that have been targeted in past CO searches. The number of detections drops at ages $\ge$10 Myr, i.e. after the transition from the protoplanetary to the debris stage of evolution \citep{Wyattsubm}, as we would expect if gas were to quickly dissipate during this transition. However, there are some sources that would now appear as exceptions: 49 Ceti and HD21997, at ages of 40 Myr \citep{Zuckerman2012} and 30 Myr \citep{Kospal2013}, respectively. These were detected in the pre-ALMA era. On the other hand, detection of CO in $\beta$ Pic has only been possible through ALMA, raising the important question of whether the presence of gas in debris systems (particularly young ones) is common or is indeed to be considered an exception. In the future, more ALMA detections (or much deeper upper limits) will be achieved for other debris discs, allowing us to address this question and to proceed with characterisation of the discs (see Sect. 5.6). The magnitude of the improvement that ALMA can provide is highlighted by our Fomalhaut upper limit lying as low as two orders of magnitude below previous upper limits.

\subsection{Characterisation with ALMA: the importance of the physical environment}

Once CO detection is reached, care needs to be taken in interpreting the observed flux. In fact, the results in Fig. \ref{fig:4} show that, in the case of Fomalhaut, a standard LTE analysis would have led to a possible underestimation of our CO mass upper limit of more than 3, and up to almost 4 orders of magnitude. This is of particular importance considering that CO detections in debris discs have been interpreted assuming LTE \citep[e.g.][]{Kospal2013,Dent2014}. Unfortunately, differences in the structure and physical conditions among different discs do not allow us to make specific predictions on deviations from LTE behaviour based on our Fomalhaut analysis; detailed modelling is indeed needed on a system-specific basis. There have been cases in the past where NLTE analysis has been carried out \citep[e.g.,][]{Hughes2008,Moor2011}; however, a priori assumptions that have been taken are more suitable for more massive, gas-rich protoplanetary discs as opposed to debris discs. For example, assuming the kinetic temperature of the gas to be equal to the dust temperature has been shown not to hold in previous modelling of optically thin discs \citep{KampvanZadelhoff2001}. In addition to this, the gas to dust ratio in debris discs is expected to be much lower than 100, considering that most gas is expected to have been depleted towards the end of the protoplanetary stage of evolution.

We have shown that in general the strong dependency of CO mass estimates on collisional partner density, kinetic temperature and radiation field cannot be neglected. This logically makes it difficult to constrain the CO content of debris discs via observations of a single optically thin line. The optimal approach would be to observe several optically thin transitions, and compare observed line ratios to expected line ratios in the $T_k$-$f_{cp}$ parameter space. 
\begin{figure}
 \vspace{-3.5mm}
 \hspace{-8mm}
  \includegraphics*[scale=0.54]{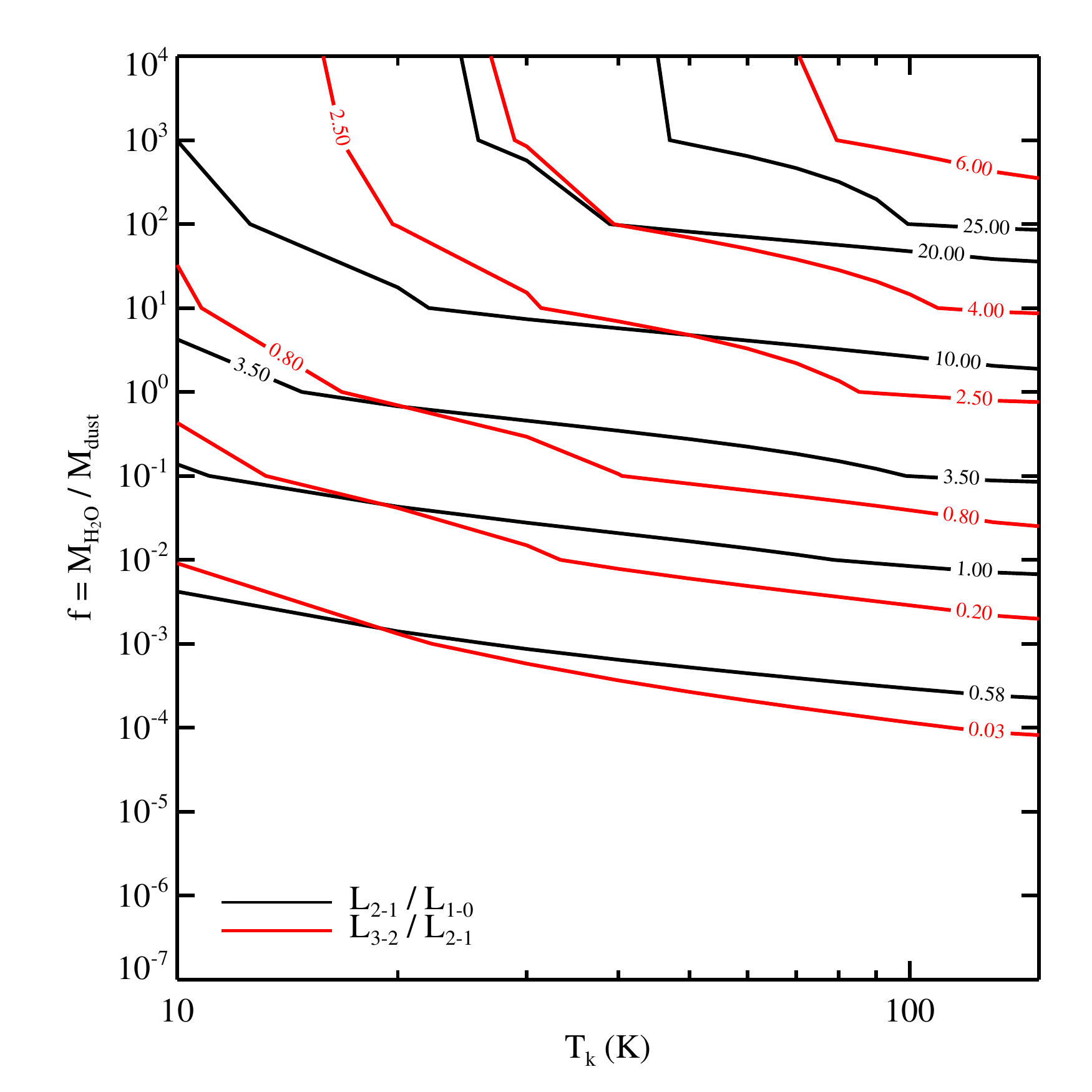}
\vspace{-8 mm}
  \caption{Water vapour-to-dust mass ratio $f_{cp}$ as a function of the kinetic temperature $T_k$ for a selection of CO line ratios. Black lines represent $L_{2-1}$/$L_{1-0}$ ratios, whereas red lines represent the $L_{3-2}$/$L_{2-1}$ ratios.}
\label{fig:8}
\end{figure}
We illustrate this concept in Fig. \ref{fig:8} for the case of the Fomalhaut ring. The procedure is as follows: we use Eq. 1 to calculate $L_{1-0}$, $L_{2-1}$ and $L_{3-2}$ for all collisional partner densities (in our qualitative example, choosing H$_2$O as the main collisional partner) and temperatures in the ranges considered in Sect. 5. The line ratios $L_{2-1}$/$L_{1-0}$ and $L_{3-2}$/$L_{2-1}$ are then computed. Assuming the far-infrared to millimetre radiation field to be well characterised, one line ratio will follow a line in $T_k$-$f_{cp}$ space (see Fig.\ref{fig:8}). This in turn means that the CO mass will be dependent on a combination of $T_k$ and $f_{cp}$ described by this line, already setting important constraints on the CO mass. In bright discs, three optically thin lines may be detected (in our example, the 1-0, 2-1 and 3-2 transitions); the two line ratios will lead to two lines in $T_k$-$f_{cp}$ space, which will cross at a single point and hence yield both the kinetic temperature of the gas and the density of collisional partners (again, see Fig.\ref{fig:8}). 

This stands for the case of optically thin emission, as it follows directly from the normalisation condition $\sum\limits^{j_{max}}_{j=0}x_{j}=1$. In the case of optically thick emission such analysis would not be appropriate, as lines would either not intercept, or intercept and lead to the wrong conclusions if optically thin emission is assumed. In order to test the optical thickness of the lines, it would be necessary to observe emission from the less abundant $^{13}$CO and C$^{18}$O isotopologues and compare it to the $^{12}$CO emission, such as was done in \citet{Kospal2013}. In the scenario where the $^{12}$CO lines are optically thick, optically thin $^{13}$CO or C$^{18}$O line ratios could be used to carry out the analysis described above in a completely analogous way.

Since ALMA delivers spatial information through images of both sub-mm continuum and line emission, this procedure would allow creation of a map of gas kinetic temperature, density of collisional partners and CO mass in the disc, gaining unprecedented insights in the physical processes at work within gas-bearing debris discs, and potentially setting tighter constraints on the origin of gas in these evolved systems.

\section{Summary and Conclusions}

In this work, we derived upper limits on the CO J=3-2 line emission in the Fomalhaut debris ring using archival ALMA Cycle-0 data, and analysed rotational excitation of the CO molecule in a debris disc environment. We achieve the following conclusions: 

\begin{itemize}

\item \textit{CO upper limits in the Fomalhaut ring: NLTE matters} \\ 
CO mass estimates derived from line fluxes in debris discs are strongly dependent on the density of collisional partners $n$, the kinetic temperature of the gas $T_k$ and the radiation field $J$ at the frequency of the transition. The most conservative upper limit on the CO mass in the Fomalhaut ring is found to be 4.9 $\times$ 10$^{-4}$ M$_{\oplus}$, while a simple LTE analysis underestimates this by $\sim$3-4 orders magnitude. \\

\item \textit{Primordial origin} \\ 
If we assume any CO in the Fomalhaut ring to be of primordial origin, H$_2$ would be its main collisional partner. Our results then show that either of, or both the CO/H$_2$ and the gas/dust ratios are much lower than their typical ISM value. In addition, the short CO photodissociation timescale compared to the age of the Fomalhaut system make this origin highly unlikely. \\

\item \textit{Secondary origin: improved upper limits} \\ 
In a more realistic picture, any CO present in the system would originate from comets, in which case we assume that H$_2$O and $\mathrm{e^-}$ are the dominant colliders. If CO/H$_2$O ratios are similar to Solar System comets, we set upper limits of 2.5 $\times$ $10^{-6}$ M$_{\oplus}$ on the CO mass, and of 1.21-0.08 $\times$ $10^{-6}$ M$_{\oplus}$ on the water vapour mass within the ring. \\

\item \textit{Secondary origin: inferring cometary compositions} \\ 
Assuming steady state, CO production/destruction rates can be used to infer the ice composition of planetesimals in a debris disc. In the Fomalhaut ring, we set an upper limit of $\sim$55\% on the CO/H$_2$O ice ratio in planetesimals. However, if LTE was to apply, this upper limit could be as low as $\sim$3\%, probing towards the low end of the range observed in Solar System comets (0.4-30\%) \\

\item \textit{CO or not CO: a matter of age or dust collision rates?} \\ 
Despite the similar dust contents of the Fomalhaut and $\beta$ Pictoris systems, the CO line emission in Fomalhaut is lower by at least three orders of magnitude. A reason may be the difference in age, since we might expect ices to have been mostly stripped off from planetesimals (via photodesorption and collisions) by the age of the Fomalhaut system (440 Myr), while being retained by the age of $\beta$ Pic (21 Myr). A second reason may be a difference in the rate of collisions within the two discs; particularly, enhanced collisional rates are expected within the clump observed in $\beta$ Pic. \\

\item \textit{Prospects for future detection with ALMA} \\  
ALMA is able to detect considerably lower amounts of CO in debris discs as compared to its predecessors, potentially revolutionising the way we think about debris discs in extrasolar planetary systems. In the Fomalhaut ring, the full ALMA array could detect CO masses down to the $\sim$10$^{-7}$ M$_{\oplus}$ level, equivalent to the CO content of $\sim$10$^4$ comets Halley. In other discs, improvements of up to several order of magnitudes on the current limits are achievable: this will allow us to understand whether debris discs contain substantial amounts of gas and how they are affected by it. In low gas density discs, we show that low transitions are best suited for detection. \\

\item \textit{Prospects for future characterisation with ALMA} \\  
We have shown that the dependence of CO mass estimates on the density of collisional partners $n$ and the kinetic temperature of the gas $T_k$ cannot be neglected. We suggest the ideal strategy for constraining the free parameters to be observation of several optically thin CO millimetre and sub-millimetre lines using ALMA. This would also allow mapping of the temperature and density structure of the gas within the disc, thus representing a considerable step forward towards understanding the origin of gas in debris systems. \\

\end{itemize}

\section*{Acknowledgements}

The authors would like to thank G. Kennedy, P. Caselli and M. Hogerheijde for useful discussions. This paper makes use of the following ALMA data: ADS/JAO.ALMA\#2011.0.00191.S. ALMA is a partnership of ESO (representing its member states), NSF (USA) and NINS (Japan), together with NRC (Canada) and NSC and ASIAA (Taiwan), in cooperation with the Republic of Chile. The Joint ALMA Observatory is operated by ESO, AUI/NRAO and NAOJ. LM acknowledges support by STFC and ESO through graduate studentships and, together with OP and MCW, by the European Union through ERC grant number 279973.

\bibliographystyle{mn2e}
\bibliography{/Users/lucamatra/Documents/lib}

\end{document}